\documentclass[12pt]{elsart}

\usepackage{epsfig}

\def\vev#1{\langle #1 \rangle}

\newcommand{\be}{\begin{equation}}
\newcommand{\ee}{\end{equation}}
\newcommand{\bea}{\begin{eqnarray}}
\newcommand{\eea}{\end{eqnarray}}
\newcommand{\bean}{\begin{eqnarray*}}
\newcommand{\eean}{\end{eqnarray*}}

\newcommand{\basispl}{
   \put(-.5,-.5){\line(1,0){1}}
   \put(.5,-.5){\line(0,1){1}}
   \put(.5,.5){\line(-1,0){1}}
   \put(-.5,.5){\line(0,-1){1}}}
\newcommand{\Upl}{\setlength{\unitlength}{.3cm}\raisebox{-.05cm}{
   \begin{picture}(1.2,1.2)(-.6,-.6)
   \basispl
   \end{picture}}}
\newcommand{\plaq}{\setlength{\unitlength}{.5cm}\raisebox{-.2cm}{
   \begin{picture}(1.2,1.2)(-.6,-.6)
   \basispl
   \put(-.5,-.5){\circle*{.2}}
   \put(-.5,.5){\circle*{.2}}
   \put(.5,-.5){\circle*{.2}}
   \put(.5,.5){\circle*{.2}}
   \put(.5,0){\vector(0,1){0}}
   \put(-.6,-.6){\makebox(0,0)[tr]{\footnotesize $x$}}
   \put(-.55,0){\makebox(0,0)[r]{\footnotesize $\nu$}}
   \put(0,-.60){\makebox(0,0)[t]{\footnotesize $\mu$}}
   \end{picture}}}
\newcommand{\onetwoplaq}{\setlength{\unitlength}{1cm}\raisebox{-.5cm}{
   \begin{picture}(1.2,1.2)(-.6,-.6)
   \put(.25,-.5){\line(0,1){1}}
   \put(.25,.5){\line(-1,0){.5}}
   \put(-.25,.5){\line(0,-1){1}}
   \put(.2,0){\line(1,0){.05}}
   \put(-.25,-.5){\circle*{.1}}
   \put(-.25,.5){\circle*{.1}}
   \put(.25,-.5){\circle*{.1}}
   \put(.25,.5){\circle*{.1}}
   \put(-.25,0){\circle*{.1}}
   \put(.25,0){\circle*{.1}}
   \put(.25,-.15){\vector(0,1){0}}
   \put(-.25,-.5){\line(1,0){.5}}
   \put(-.3,-.55){\makebox(0,0)[tr]{\footnotesize $x$}}
   \put(-.3,0){\makebox(0,0)[r]{\footnotesize $\nu$}}
   \put(0,.8){\makebox(0,0)[t]{\footnotesize $\mu$}}
   \end{picture}}}
\newcommand{\twooneplaq}{\setlength{\unitlength}{.5cm}
   \raisebox{-.2cm}{
   \begin{picture}(2.2,1.2)(-1.1,-.6)
   \put(-1,-.5){\line(1,0){2}}
   \put(-1,.5){\line(1,0){2}}
   \put(-1,-.5){\line(0,1){1}}
   \put(1,-.5){\line(0,1){1}}
   \multiput(-1,-.5)(1,0){3}{\circle*{.2}}
   \multiput(-1,.5)(1,0){3}{\circle*{.2}}
   \put(-1.1,-.6){\makebox(0,0)[tr]{\footnotesize $x$}}
   \put(-1.05,0){\makebox(0,0)[r]{\footnotesize $\nu$}}
   \put(-.45,-.60){\makebox(0,0)[t]{\footnotesize $\mu$}}
   \put(1,0){\vector(0,1){0}}
   \end{picture}}}

\begin{document}

\begin{frontmatter}
\title{Heavy Quarkonia at High Temperature}
\author{Jochen Fingberg}
\address{Department of Physics,
         P.O. Box 10~01~27,
         University of Wuppertal,\\
         42097 Wuppertal,
         Germany}
 
\begin{abstract}
We present a new method to study the properties of heavy quarks
at finite temperature. It combines non-relativistic QCD
with an improved gluonic action on anisotropic lattices.
The efficiency of the approach is demonstrated
by the first non-perturbative calculation of
the temperature dependence of low-lying quarkonium "pole" masses.
For ground state meson masses
in the region between charmonium and bottomonium
we find only very little variation
up to our highest temperature which corresponds to $T\approx 1.2~T_c$
while first excited states indicate a large mass shift.

\end{abstract}
\end{frontmatter}

\section{Introduction}
At high temperature hadronic matter is expected to undergo a
phase transition to the quark-gluon plasma (QGP).
In recent years there have been significant theoretical
and experimental developments to study it's properties.
High energy heavy-ion experiments designed to detect spectral changes 
of hadrons in hot media have already been started.
In the search for possible signals of the QGP heavy quarkonium states are among
the simplest probes that allow to test the structure of the QCD vacuum.
For instance $J/\Psi$-suppression due to colour screening
has been proposed to probe deconfinement \cite{MatsuiSatz}.
Quarkonium production is rather well understood
in hadron-hadron collisions \cite{qprod}. However,
the behaviour of bound states of heavy quarks
in a strongly interacting medium close to the deconfinement
temperature $T_c$ is still largely uncertain.

Previous theoretical results on the temperature dependence of the masses
of the $\eta_c$, the $J/\Psi$, and the $\Psi^\prime$
are not yet completely satisfactory.
Different calculations give model dependent results.
Some predict a significant decrease \cite{pot,pot1,precursory,sum_rules}
while others suggest that the masses stay constant \cite{NJL}
or even rise with temperature \cite{NaNu,ftmes93,ftmes95}.
The assumptions for instance about a temperature dependence of the
string tension, $\sigma(T)$, the strong coupling constant, $\alpha_s(T)$,
the effective quark mass $M_Q(T)$ and the gluon condensate 
 $\langle \Omega | F_{\mu\nu} F^{\mu\nu} | \Omega \rangle = G^2(T)$
primarily depend on perturbation theory and may not be reliable
in the temperature region under consideration.

Progress in numerical simulations of QCD
makes reliable predictions about hadronic properties
in the non-perturbative regime possible.
Recently a non-relativistic approximation of QCD has been used to
reproduce the experimental spectrum for
heavy quarkonia with high precision \cite{CTHD94,CTHD95}.

At finite temperature a complication
arises because Lorentz invariance is explicitly
broken so that Green's functions defined
by correlators in Euclidean time and space directions are
controlled by different phenomena. The heavy quark potential which is
confining in the low temperature region becomes Debye-screened at high
temperature. On the other hand the pseudo-potential from
spatial Wilson loops is confining
for all temperatures \cite{MaPo87,SST93,KLL95}.
So far most of the work has been concentrated
on the calculation of screening masses obtained from spatial
correlators which do not have
a direct connection to the physical mass of a resonance
as defined for instance by the position of a peak in the spectral function.
In recent investigations of the temperature dependent structure in
the light mesonic channels screening masses from spatial correlations
have been found to differ from effective pole masses from correlations in the 
Euclidean time direction \cite{ftmes93,ftmes95}.
The mass shift of hadrons made of light quarks
is expected to be mainly controlled by chiral symmetry restoration.
A realistic simulation would require the inclusion of light dynamical
fermions which is expensive. To begin with,
it is advantageous to study the temperature dependence
of the spectrum with heavy quarks.
In this case the influence of light quarks
is less important so that a quenched simulation
is a reasonable approximation.
The binding energy for $c\bar c$ ($\approx 0.63$~GeV)
and $b\bar b$ ($\approx 1.1$~GeV) \cite{PDG} is large
compared to the deconfinement
temperature $T_c \approx 150-250$~MeV.
At intermediate temperatures the spectral width of the
low lying $q\bar q$ bound states is expected to be small
so that there will still be a clear distinction
between the continuum and the lowest resonance.
The binding energy of quarkonia decreases with temperature.
The dissociation temperature of the $(\Upsilon,~\Upsilon^\prime)$
is expected to be in the region of $T\approx (2.6,~1.1)~T_c$ \cite{Tdiss}.
However, these values still contain a model dependent uncertainty.

Our approach is based on an improved gluonic action for the
light degrees of freedom and
a non-relativistic formalism (NRQCD) for the heavy quarks \cite{Lepage}.
Finite temperature NRQCD (FT-NRQCD)
uses anisotropic lattices \cite{asym,aniso,colin} to achieve
a finer resolution in time direction.
A large number of Matsubara frequencies is necessary
to accurately measure temporal meson propagators.
The problem is first approached in quenched QCD
by considering a quark-antiquark pair propagating in Euclidean time
direction in a gluonic medium. To show the feasibility
of the new approach and the significance of its results
we calculate temporal meson correlators
for bare quark masses between 2 and 6~GeV.

\section{Heavy meson spectrum at finite temperature}
Heavy quarkonia are small and tightly bound. 
We know that asymptotically for infinite quark mass a
potential picture will give the correct description.
The Cornell potential
\be
    V(r) = \sigma r - \frac{\alpha_s}{r}
\ee
reproduces the experimental spectrum for charmonium and bottomonium quite well.
In a thermal medium of temperature $T>0$ the potential is modified
by colour screening which can be parameterized in the form \cite{Montvay}
\be
    V_T(r) = \frac{\sigma}{\mu(T)} (1-\exp(-\mu(T)r)) -
             \frac{\alpha_s}{r} \exp(-\mu(T)r) ~~~~~~.
\label{eq:screened}
\ee
Eq.~\ref{eq:screened} is equivalent to the Cornell form with
a temperature dependent string tension
 $\sigma(T)=\sigma/(\mu(T)r)\left(1-\exp(-\mu(T)r)\right)$
and a screened Coulomb term $\alpha_s(T)=\alpha_s\exp(-\mu(T)r)$.
\begin{figure}[htb]
\hbox{
\epsfig{file=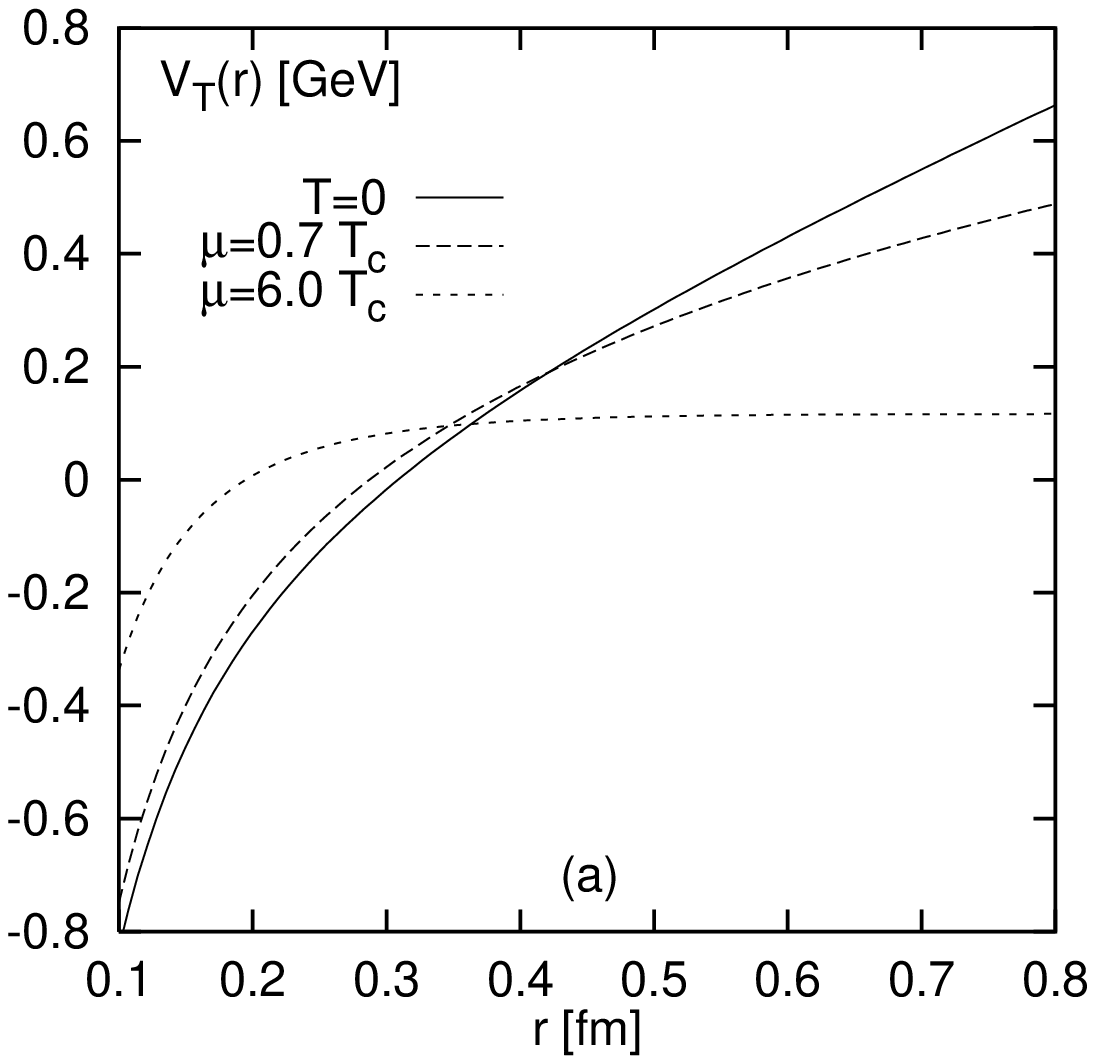,bbllx=83,bblly=48,
        bburx=396,bbury=353,clip=,width=0.325\linewidth}
\epsfig{file=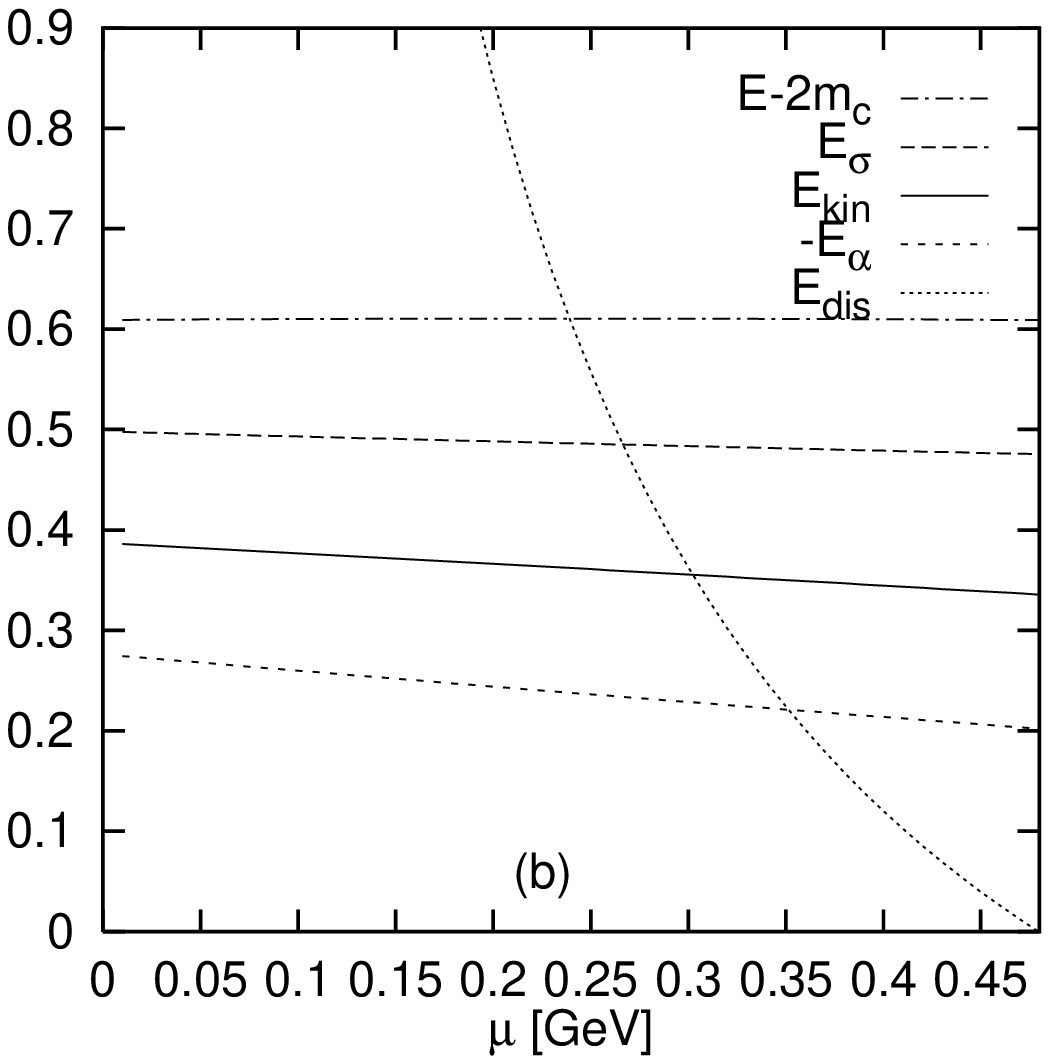,bbllx=83,bblly=48,
        bburx=388,bbury=353,clip=,width=0.32\linewidth}
\epsfig{file=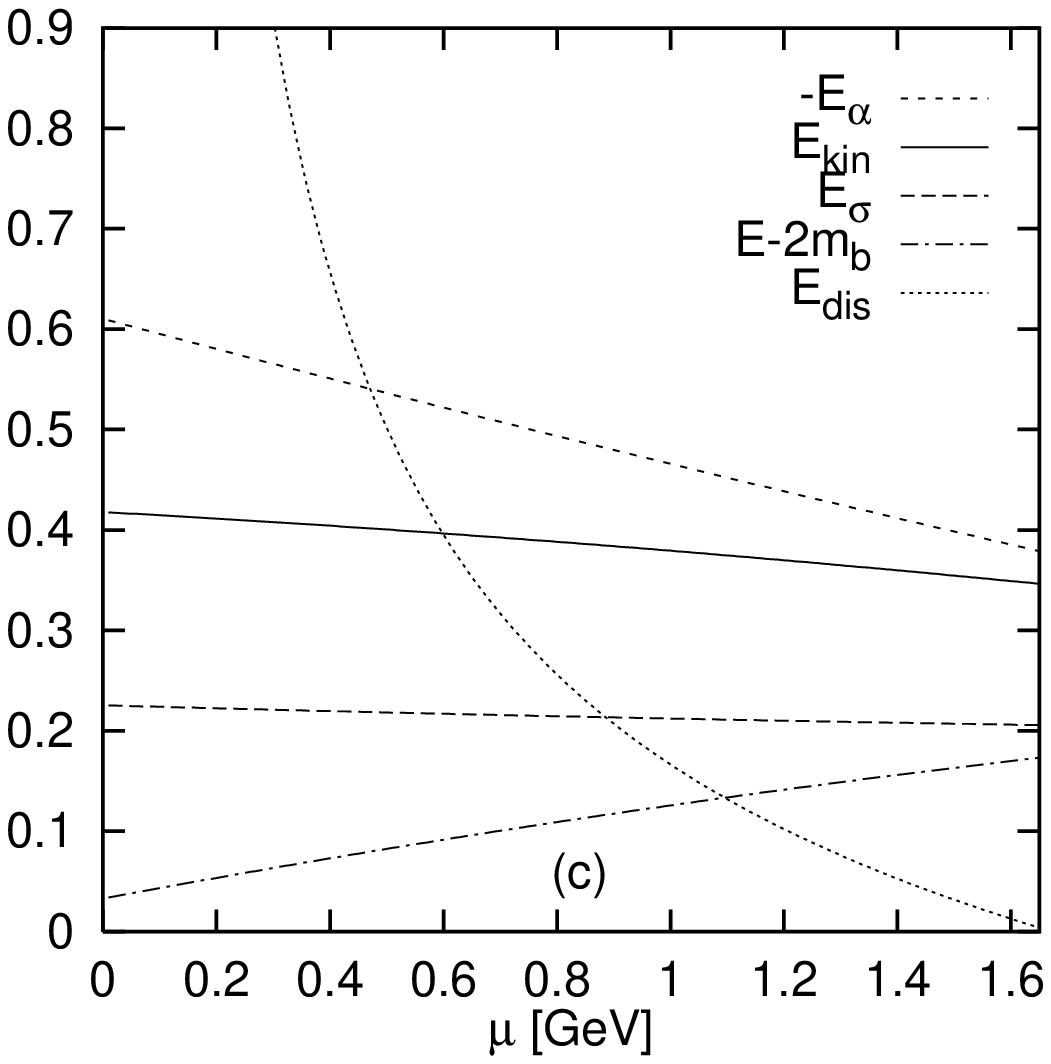,bbllx=83,bblly=48,
        bburx=388,bbury=353,clip=,width=0.32\linewidth}
     }
\caption{The heavy quark potential (a)
         as a function of the radius
         for 3 different temperatures
         and the terms contributing to the
         total energy together with the dissociation
         energy in GeV as a function of the Debye-mass
         for $c\bar c$ (b) and $b\bar b$ (c).}
\label{fig:pot_v}
\end{figure}
From fig. \ref{fig:pot_v}a we see that $V_T(r)$ changes differently
with temperature for small $r<0.3~$fm and large $r>0.4$~fm.
The temperature dependence is entirely contained in the Debye-mass.
It has been argued \cite{Satz} that due to string breaking $\mu(T)$
will be different from zero even for $T=0$ and that the
effective string tension does not vanish immediately above $T_c$
modeling non-perturbative interactions in the plasma. Although
the precise functional dependence is not very well known
the string tension and the strong coupling will decrease
when the temperature and thereby $\mu(T)$ is increased.

First qualitative insight in the behaviour of $Q\bar Q$-states can be
obtained from a semi-classical picture \cite{Satz}.
As a consequence of the uncertainty relation
which forces $\vev{p^2}\vev{r^2}\approx 1$ the kinetic energy,
$E_{kin}=p^2/m\approx 1/mr^2$, decreases with the quark separation.
Minimizing the total Energy, $E(r)=2m+E_{kin}+V$, the ground state
is found where the decrease of the kinetic energy is compensated
by an increase of the potential energy. The result is that
the binding radius $r_0(\mu)$ increases with $\mu(T)$.
A minimum of $E(r)$ exists as long as the screening is not too strong.
The meson will dissociate once the Debye-mass
$\mu(T)$ becomes larger than a critical value $\mu_c$.
Below this value the meson mass depends on the temperature.
In general the sign and the magnitude of the mass shift,
$\Delta M=M(T)-M(T=0)$, depend on the details of the
balance of potential (Coulomb and string) and kinetic energy
as is shown in figs.~\ref{fig:pot_v}b and c.
In the potential of eq.~\ref{eq:screened} masses
can decrease or increase with $T$
depending on the size of the $Q\bar Q$ bound state.

Understanding the basic mechanism in the semi-classical approximation
we can move towards a quantitative understanding and
calculate quarkonium wave-functions in
a non-relativistic potential model.
The mean quark velocity can be computed from averages
obtained by a numerical solution of the Schr\"odinger equation
with a potential of the form given in eq.~\ref{eq:screened},
\be
 \vev{v^2}_T = \frac{E(T)-\sigma(T) \vev{r}_T +\alpha_s(T)\vev{1/r}_T}{M}
 ~~~~.
\ee
\begin{figure}[htb]
\hbox{
\epsfig{file=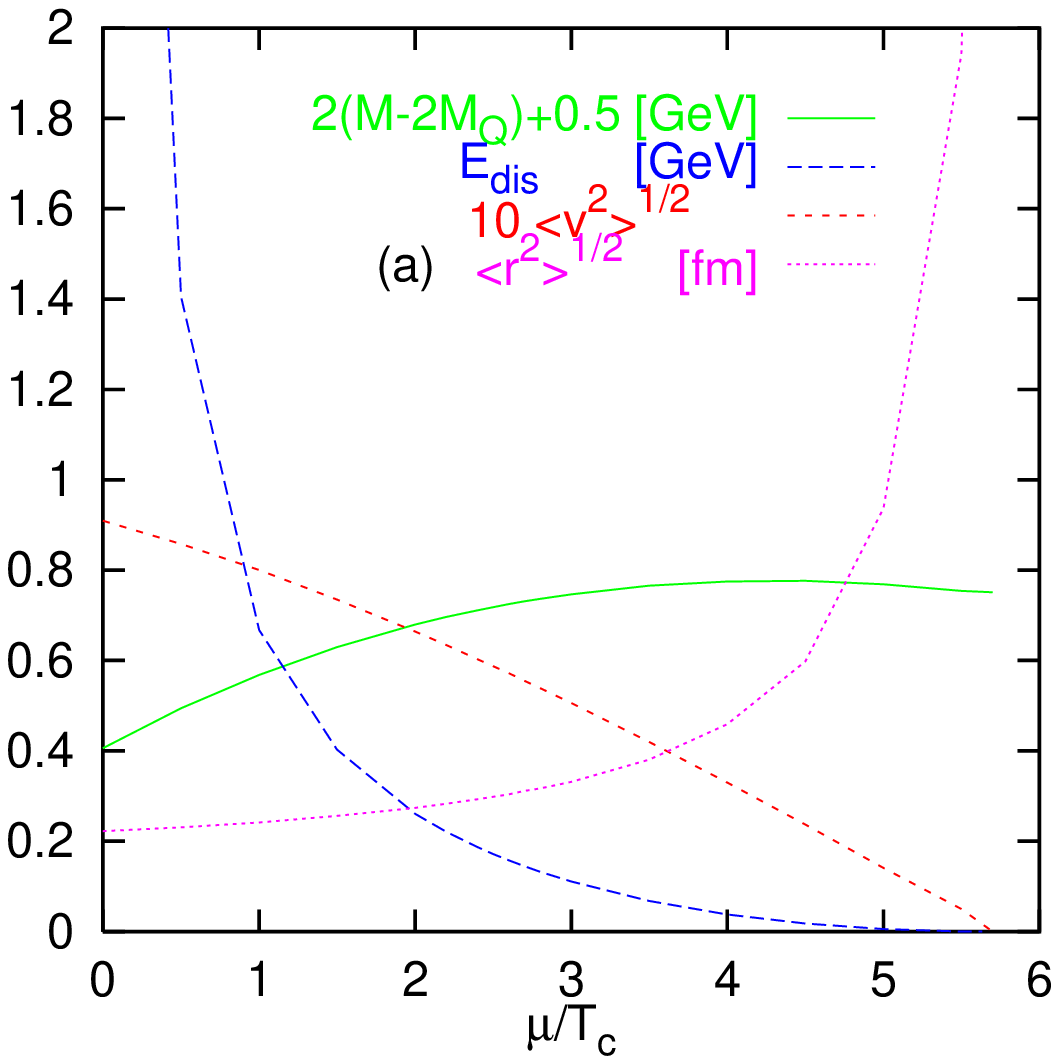,bbllx=88,bblly=48,
        bburx=391,bbury=353,clip=,width=0.48\linewidth}
\epsfig{file=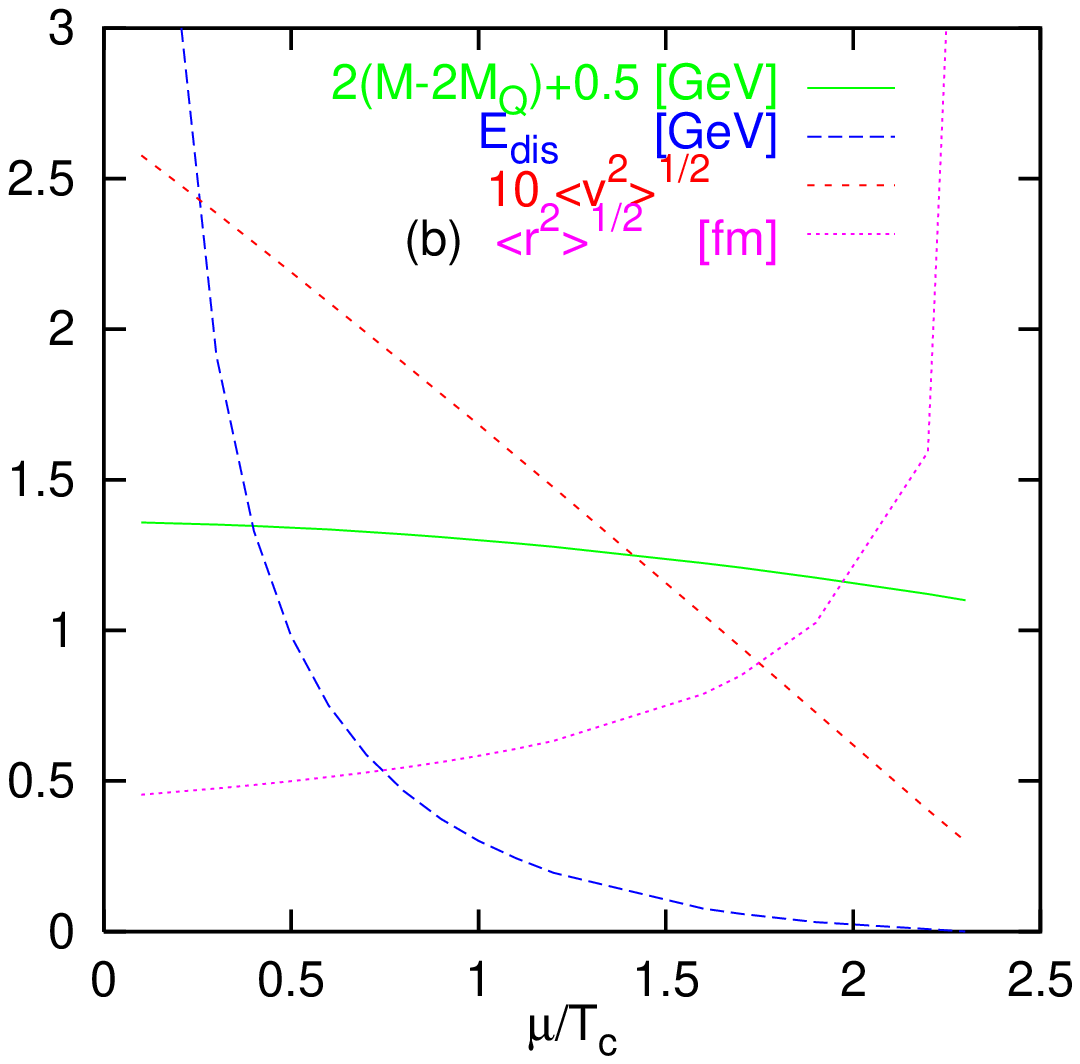,bbllx=88,bblly=48,
        bburx=395,bbury=353,clip=,width=0.48\linewidth}
     }
\caption{Dependence of the velocity, radius, mass and dissociation energy
         in the (1S) state on the Debye-mass for $b\bar b$ (a) and
         $c\bar c$ (b).
         The velocity and the meson mass have been rescaled
         to fit in the same frame.
}
\label{fig:vrme}
\end{figure}

From fig.~\ref{fig:vrme} we see that the average quark velocity
decreases with temperature so that a non-relativistic ansatz
for the heavy degrees of freedom seems to be justified for $T>0$.

However, beyond this simple potential picture
we know that lattice data do not agree very well
with the perturbative expectation
for the form of the heavy quark potential at least
up to $T \le 4~T_c$ \cite{Edwin,Bengt,columbia}.
A precise measurement of the Debye-mass is plagued by ambiguities
in the form of the function used to fit the heavy quark potential.

Lattice simulations can provide a reliable determination
of the meson spectrum without assumptions about the heavy quark potential.
Thus, it is important to compare phenomenological and numerical results
to gain further insight in the structure of the QCD vacuum at finite
temperature.

Compared to zero temperature additional care
is necessary because any excitation
acquires a finite lifetime at non-zero temperature \cite{Henning}.
The spectral function $\rho(p,\omega)$
of an excitation with momentum $p$ and energy $\omega$
which is a $\delta$-function at $T=0$ will broaden.
This effect will modify the meson propagator, $G_m(t)$.

\section{FT-NRQCD}
We can go beyond a potential model approximation and calculate
the action of heavy quarks in a gluonic heat-bath from first principles.
The proposed method is based on the formalism of non-relativistic
QCD \cite{Lepage} and a discretization of
quantum field theories at finite temperature on
anisotropic lattices \cite{asym,aniso,colin}.
NRQCD is an effective theory that
has been successfully applied to bottomonium and charmonium
at zero temperature \cite{CTHD94,CTHD95}.
It allows an efficient and accurate calculation of 
heavy quark propagators. At zero temperature
the binding energy of a $b\bar b$-system can be estimated 
from it's distance to the open $b$ threshold,
$E_b=M_{b\bar b}-2m_B\approx 1.1$~GeV.
The binding energy is large compared to the
deconfinement temperature, $T_c \approx 150-250$~MeV,
so that we expect a smooth evolution of the system away from it's
ground state at $T=0$ as the temperature is increased.
From previous considerations of potential models
we expect the average quark velocity to decrease with temperature
so that there will be a range of temperatures $T>0$
where FT-NRQCD is applicable.

The total action for a system of heavy quarks
in a gluonic heat-bath naturally splits into a
relativistic part for the gluons
and a non-relativistic term for the heavy quarks.
\be
       S = S_R +S_{NR}
\ee
The relativistic simulations use a tree-level
improved gauge action \cite{tia,anisotrop}
with a plaquette and a rectangle term on asymmetric lattices
with separate spatial coupling $\beta_\sigma$ and temporal coupling
$\beta_\tau$.
{\samepage
\bea
  \!\!\! S_R &=& \beta\left[\gamma^{-1}\sum_{x\atop{4>\nu>\mu}}\left(
                    \left(1-\plaq      \right)-
        c_\sigma    \left(1-\twooneplaq\right)-
        c_\sigma    \left(1-\onetwoplaq\right)\right)\right.\nonumber\\[0.2cm]
 +&\gamma&\sum_{x\atop{4=\nu>\mu}}\left.\left(\frac{u_\sigma^2}{u_\tau^2}
                    \left(1-\plaq      \right)-
        c_\tau      \left(1-\twooneplaq\right)-
        c_\tau\frac{u_\sigma^2}{u_\tau^2}
                    \left(1-\onetwoplaq\right)\right)\right]\\
  c_\sigma &=& \frac{1}{20~u_\sigma^2},~~~~~~
  c_\tau    =  \frac{1}{20~u_\tau^2}   \nonumber
\eea
}
The asymmetry parameter $\gamma=\sqrt{ \beta_\tau / \beta_\sigma }$ and
the coupling $\beta=\sqrt{ \beta_\tau ~ \beta_\sigma }$ are
defined in terms of the spatial coupling $\beta_\sigma$
and the temporal coupling $\beta_\tau$.
In principle, tadpole improvement can be implemented by factors
\[
u_\sigma=\langle \frac{1}{3}~\mbox{Re~Tr~}\Upl_{\sigma\sigma}\rangle^{1/4}
 ~~~~\mbox{and}~~~~
u_\tau =\langle \frac{1}{3}~\mbox{Re~Tr~}\Upl_{\sigma\tau }\rangle^{1/4}
\]
which can be calculated from the average spatial and temporal plaquette
as indicated or the mean link in Landau gauge.
The anisotropy $\xi=a_\sigma/a_\tau >1$ becomes equal
to the asymmetry parameter $\gamma$
only in the continuum limit, $\beta\rightarrow\infty$.
At finite gauge coupling, they differ
by a renormalization factor $\eta=\xi/\gamma$ which can be determined
nonperturbatively in a calibration procedure.
The additional parameter $\gamma$ allows to have a large number of lattice
points in time direction while keeping the temperature
$T=(N_\tau~a_\tau)^{-1}=(N_\tau~\xi~a_\sigma)^{-1}$ fixed.

The non-relativistic action is derived from the Dirac equation
by a Foldy-Wouthuysen transformation. The resulting
effective field theory called NRQCD approximates
relativistic QCD at small energies. Relativistic
heavy-quark momenta are excluded from the theory by
choosing a spatial lattice spacing $a_\sigma^{-1} \simeq M_Q$.
The $1/M_Q$ expansion underlying the formalism of NRQCD
will hold as long as the average quark velocity is small, $v \ll 1$.
The non-relativistic action has the form
\be
     S_{NR} = \Psi^\dagger (D_t + H_0 + \delta H) \Psi
\ee
where $D_t$ denotes the covariant
time derivative, $H_0$ is the kinetic energy operator
and $\delta H$ is the leading relativistic and
finite-lattice-spacing correction.
Here we include all spin-independent relativistic
corrections to order $M_Q v^4$ and spin-dependent corrections
to order $M_Q v^6$.
Modifications to the corresponding form
of $S_{NR}$ for $T=0$ as given in ref.~\cite{rad}
appear only in the correction $\delta H$
where the improved temporal derivative and the 
chromoelectric field strength introduce additional
factors of $\xi$. To simplify matters, tadpole improvement of $S_{NR}$
was implemented only for spatial links. In our case the spatial lattice
spacing is considerably larger than the temporal spacing so that
the mean temporal link $u_\tau$ is very close to unity \cite{colin}.

The heavy quark propagators are computed using the evolution equation
\be
  G_{t+a_\tau} = \left(1-\frac{a_\tau H_0}{2}\right)~U_4^\dagger~
                 \left(1-\frac{a_\tau H_0}{2}\right)~
                 \left(1-{a_\tau \delta H}   \right)~G_t
\ee
The non-relativistic quark propagator is not periodic in time and can
be evaluated at times larger than $N_\tau/2$.
A symmetric (antisymmetric) propagator can be constructed by explicitly
adding (subtracting) the contributions from mirror charges.
On a Euclidean lattice at temperature $T=1/(a_\tau N_\tau)$ thermal
Green's functions can be evaluated only on a discrete set of frequencies
$\omega_n=2\pi n T$, $n=1 ... N_\tau$. It is obvious that a good resolution
requires a large number of grid points in temperature direction.
Thermal meson states
\be
G_m(p,t)=\sum_x\mbox{Tr}\left[\sum_r G_t^\dagger(x-r)\Gamma^{(sk)}(r)~
                              \sum_s G_t(x-s)        \Gamma^{(sc)}(s)\right]
     ~{\mathrm e}^{ipx}
\ee
corresponding to ${}^3S_1$ and ${}^1S_0$-states were constructed from the
quark propagators. Interpolating operators of the form
$\Gamma(r) = \Omega^{\mathrm spin}~\Phi(r)$ with 11 different
combinations of local and smeared trial
wave-functions at the source and sink were used.
The wave-functions were determined solving the
Schr\"odinger equation for a Breit-Fermi potential
 \cite{BF} for each value of the bare quark mass.
The same set of wave-functions was used for all temperatures.

\section{Simulation details and results}
For the relativistic simulations we choose an asymmetry
parameter $\gamma=4$ so that our largest lattice
size, $16^3 \times 64$ corresponds to a symmetric
lattice at zero temperature.
The temperature was varied by reducing the
value of $N_\tau$ from 64 to 24 and 16
for fixed spatial lattice size, $N_\sigma=16$.
The value of the bare coupling, $\beta=4.31466$,
corresponds to the critical coupling
$\beta_{\mathrm{Wilson}}=5.8941$ \cite{Cella94,thermo}
for $\gamma=1$ on a lattice with $N_\tau^{\mathrm{critical}}=6$.
In this way it is possible to get a rough estimate of the 
spatial lattice spacing from
the string tension $\sqrt{\sigma}a=0.2734(37)$
at $\beta_{\mathrm{Wilson}}=5.8941$ \cite{thermo}.
We expect the inverse spatial lattice spacing to be
in the region $a_\sigma^{-1}=\sqrt{\sigma}/0.2734\approx 1.5$~GeV.
However, a value of $\gamma\ne 1$ will modify this correspondence
and change the value of the spatial lattice spacing.
The remaining free parameter,
the bare quark mass, was varied in the
range $a_\sigma M_Q = 1.5, 2, 2.5, 3, 3.5, 4$.
After generating $\approx 100$ independent gauge field configurations
we determined zero momentum
meson propagators for ${}^1S_0$ and ${}^3S_1$ states. For local
${}^1S_0$ states we also measured finite momentum propagators.

Before we can measure the meson spectrum
we have to determine the asymmetry $\xi=a_\sigma/a_\tau$ and
set the scale $a_\sigma$.
The heavy quark potential at $T=0$ can be measured either
from spatial or temporal Wilson loops.
The asymmetry parameter is determined in a calibration procedure
from spatial and temporal potential differences.
\begin{figure}[htb]
\hbox{
\epsfig{file=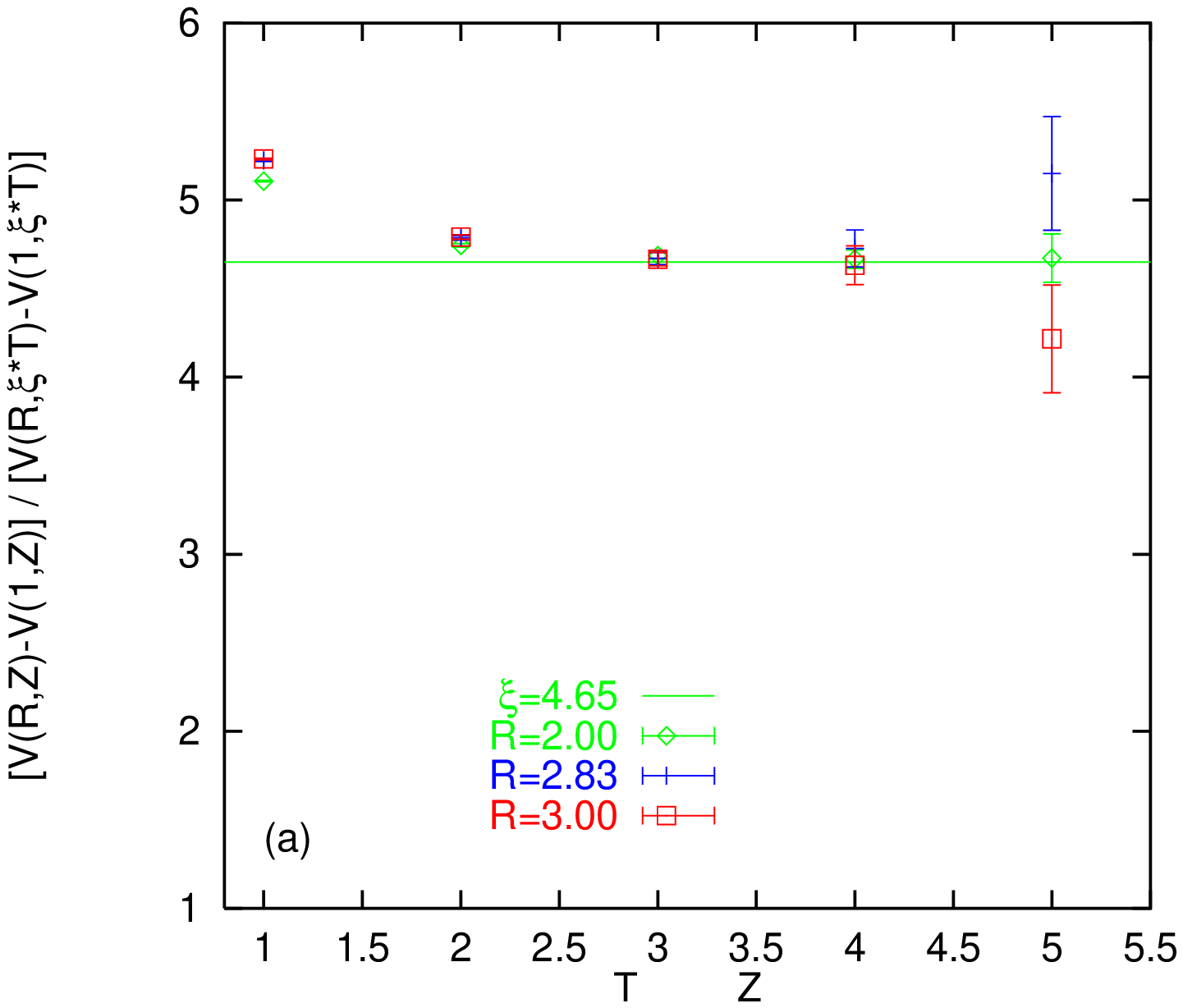,width=0.5\linewidth}
\epsfig{file=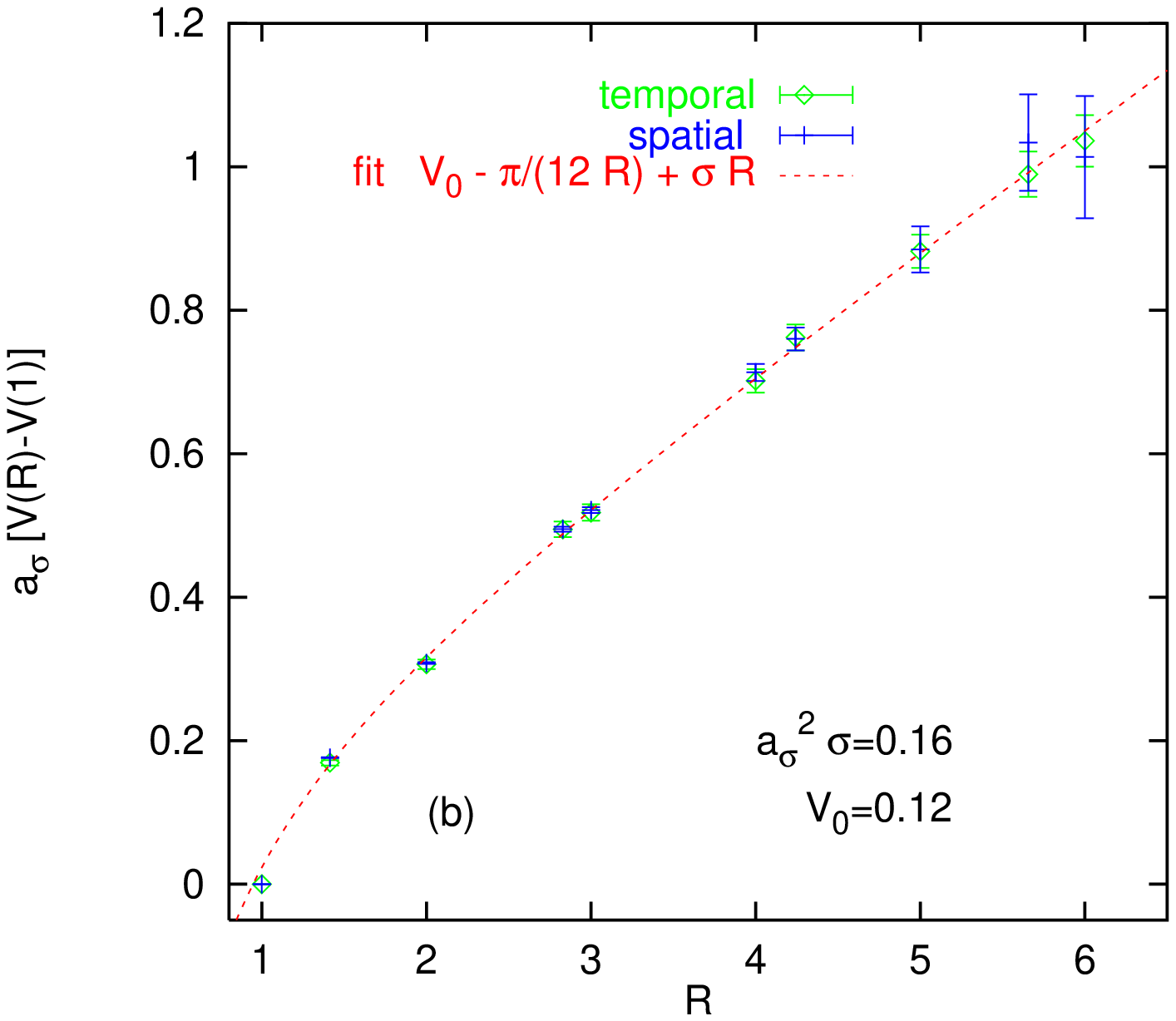,width=0.5\linewidth}
     }
\caption{The ratio of spatial and temporal potential differences (a) and
         heavy quark potential from temporal and spatial Wilson loops (b).}
\label{fig:calibration}
\end{figure}
The parameter $\xi$ can be calculated most accurately
at small distances where the
statistical error is small. In fig. \ref{fig:calibration}a we see
that for the smallest separation $R=2$ the ratio of potential
differences agrees with $\xi=4.65$ for all separations $T>2$. All values
for larger separations $R=2.83$ and $R=3$ are in accord with this value
within the statistical errors. Fig.~\ref{fig:calibration}b shows that spatial
and temporal Wilson loops give the same physical potential when this value of
the asymmetry parameter is used. Knowing $\xi$ and $a_\sigma \sqrt{\sigma}$
an estimate of the temperature can be obtained from the relation
\[
 \frac{T}{T_c}=\frac{T}{\sqrt{\sigma}}~\frac{\sqrt{\sigma}}{T_c}
      =\frac{\xi}{N_\tau (a_{\sigma}\sqrt{\sigma}) (T_c/\sqrt{\sigma})}
\approx \frac{19}{N_\tau} \approx 0.8,1.2~(N_\tau=24,16) ~~,
\]
where we use a value of $T_c/\sqrt{\sigma}=0.625$ \cite{thermo}.
A least squares fit to the static potential with the functional
form ~$V(R)=V_0-\pi/(12R)+\sigma R$~
gives the result $a_\sigma^2 \sigma =0.16$. This translates
to $a_\sigma^{-1}\approx 1.1$~GeV when the phenomenological
value for the string tension, $\sqrt{\sigma}=427$~MeV, is used.
\begin{table}
\caption{The kinetic mass and hyperfine splitting
         of the ground state
         for various values of the bare quark mass
         for $\xi=4.65$ and $\Delta E_{hyp}$
         between 15 and 25~MeV.}
 \begin{tabular}{c|c|c|c}
 $a_\sigma~M_Q$ &
 ~~$\xi a_\sigma M_{kin}$~~ &
 $a_\sigma^{-1}=
  \frac{\xi~M_{\eta_b}}{\xi a_\sigma M_{kin}}$ &
 $a_\tau \Delta E_{hyp}=\frac{\Delta E_{hyp}}{\xi~a_\sigma^{-1}}$\\ \hline
  1.5 & 15.4 (~2) & 2.85~GeV & 0.0011~~--~~0.0019  \\
  2.0 & 20.1 (~3) & 2.18~GeV & 0.0015~~--~~0.0025  \\
  2.5 & 24.8 (~4) & 1.77~GeV & 0.0018~~--~~0.0030  \\
  3.0 & 29.6 (~5) & 1.48~GeV & 0.0022~~--~~0.0036  \\
  3.5 & 34.5 (~7) & 1.27~GeV & 0.0025~~--~~0.0042  \\
  4.0 & 39.8 (10) & 1.10~GeV & 0.0029~~--~~0.0049  \\
 \end{tabular}
\label{tab:test}
\end{table}
To show the consistency a second estimate for
the spatial lattice spacing is determined
from the hyperfine splitting
of the ground state, $\Delta E_{hyp}$.
Two conditions are needed to fix the
unknown parameters $a_\sigma^{-1}$ and $M_Q$.
First we set the scale and determine $a_\sigma^{-1}$ from
$M_{kin}=M_{\eta_b}=M_\Upsilon - \Delta E_{hyp}$ where we use
the experimental value for the mass of
the $\Upsilon(1S)=9.46037(21)$~GeV \cite{PDG}.
The kinetic mass $M_{kin}$ is determined from
a fit to a non-relativistic dispersion relation,
\be
     M(p) = M_1 + \frac{|P|^2}{2M_{kin}} + ... ~~~~~~~~~
\ee
where $M(p)$ is the mass extracted from finite momentum propagators.
From previous calculations $\Delta E_{hyp}$ is expected to be in the
region between 15 and 25~MeV at our presumably
large value of $a_\sigma^{-1}$ between 1.1 and 1.5~GeV \cite{Tsukuba}.
\begin{figure}[hbt]
\begin{centering}
\epsfig{file=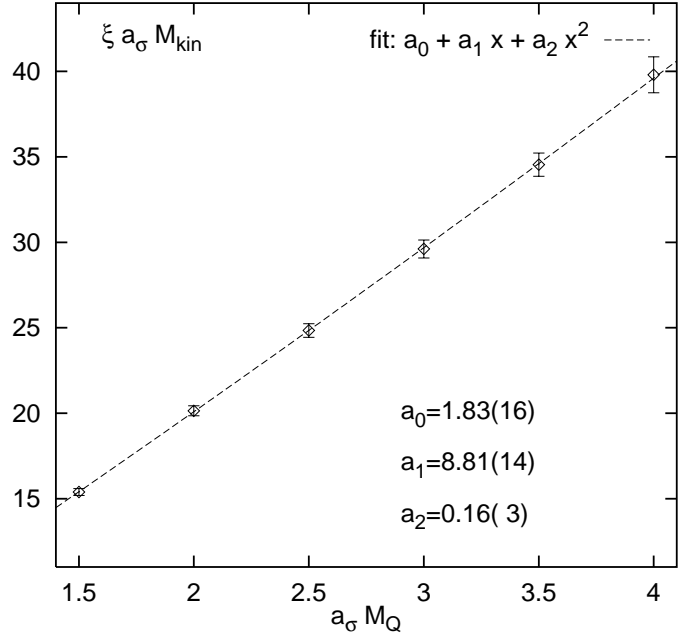,width=0.75\linewidth}
\caption{The kinetic mass of the ${}^1S_0$ meson
         as a function of the bare quark mass.}
\label{fig:Mkin}
\end{centering}
\end{figure}

\begin{figure}[htb]
\vbox{
\hbox{
\epsfig{file=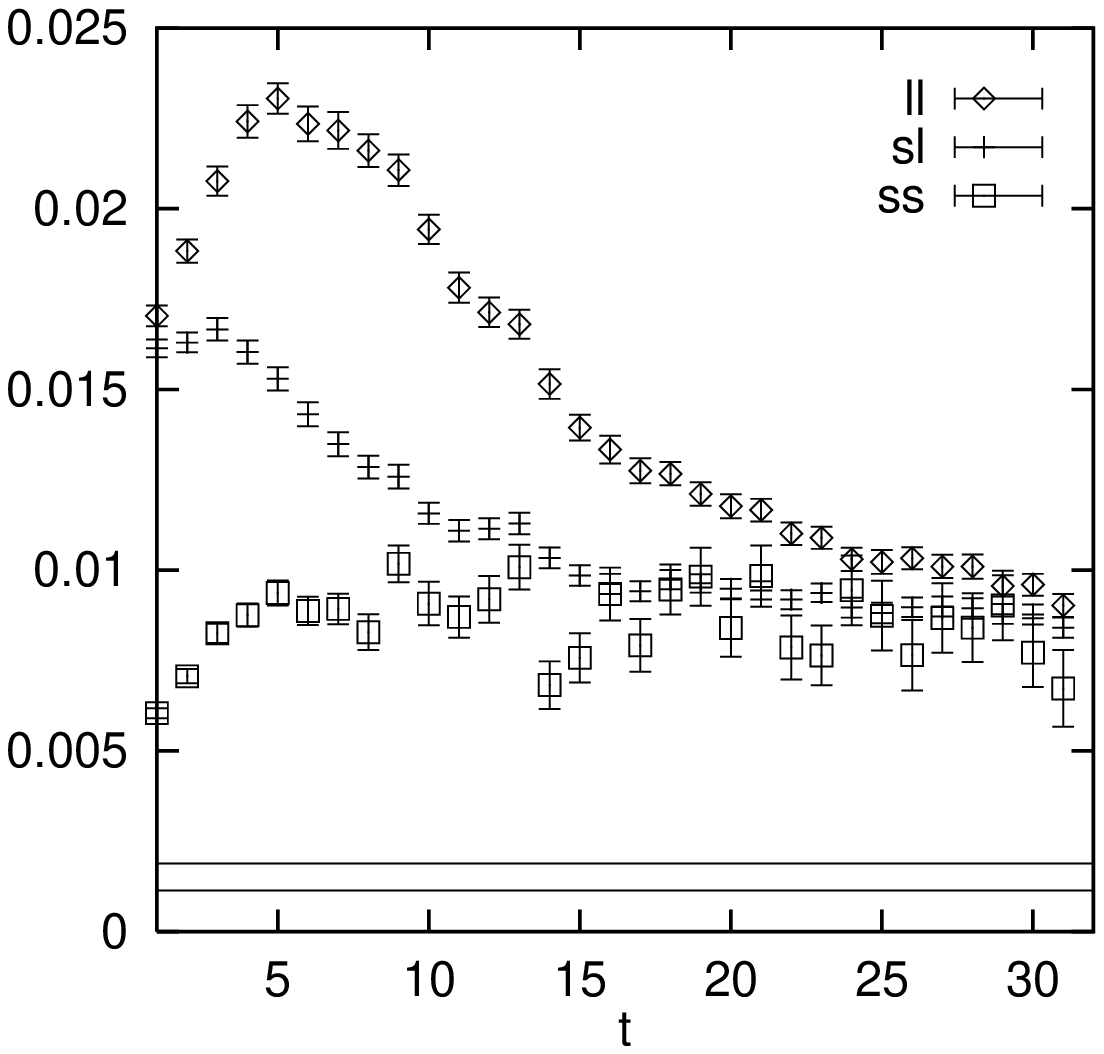,bbllx=71,bblly=52,
        bburx=388,bbury=346,width=0.481\linewidth}
\epsfig{file=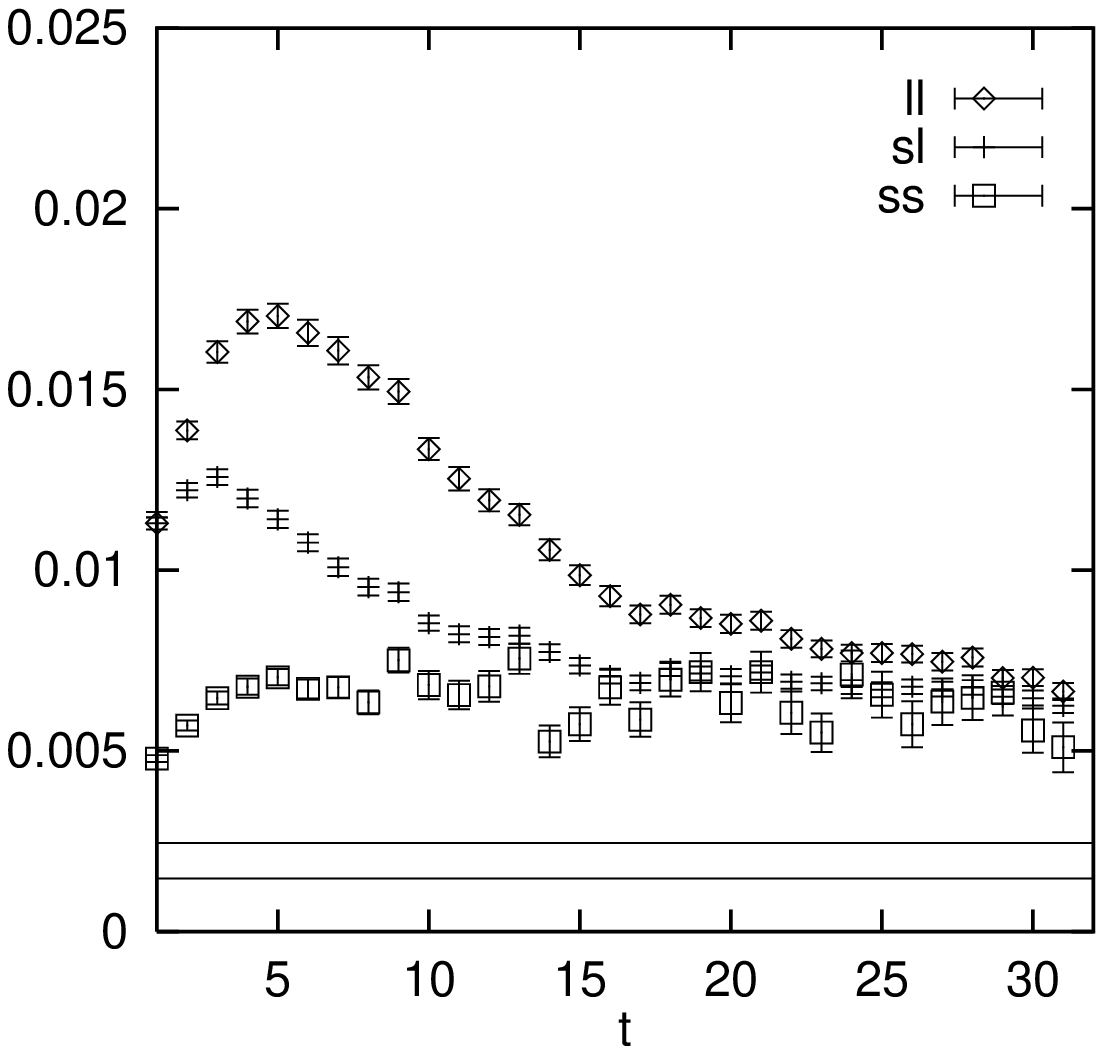,bbllx=71,bblly=52,
        bburx=388,bbury=346,width=0.481\linewidth}
     }
\hbox{
\epsfig{file=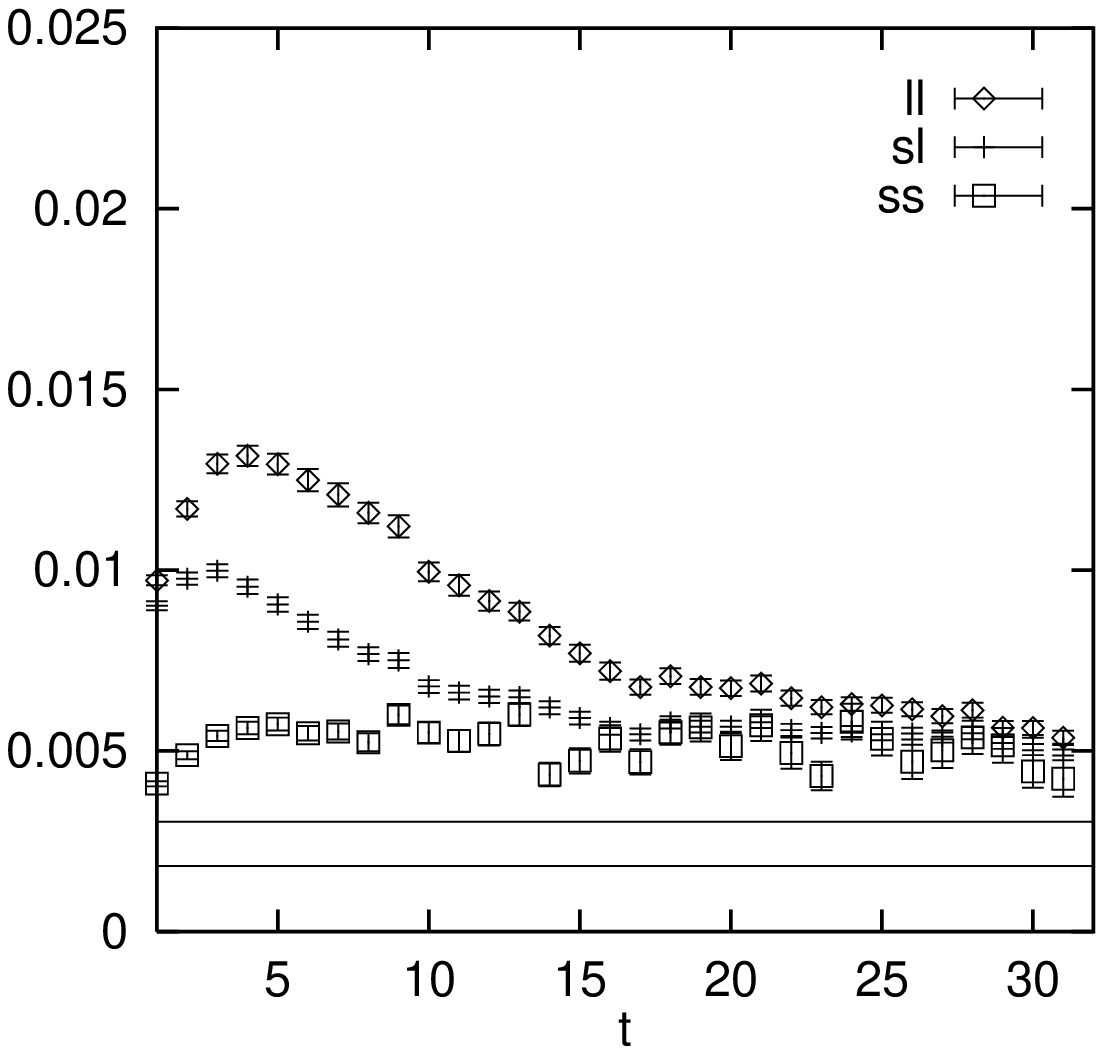,bbllx=71,bblly=52,
        bburx=388,bbury=346,width=0.481\linewidth}
\epsfig{file=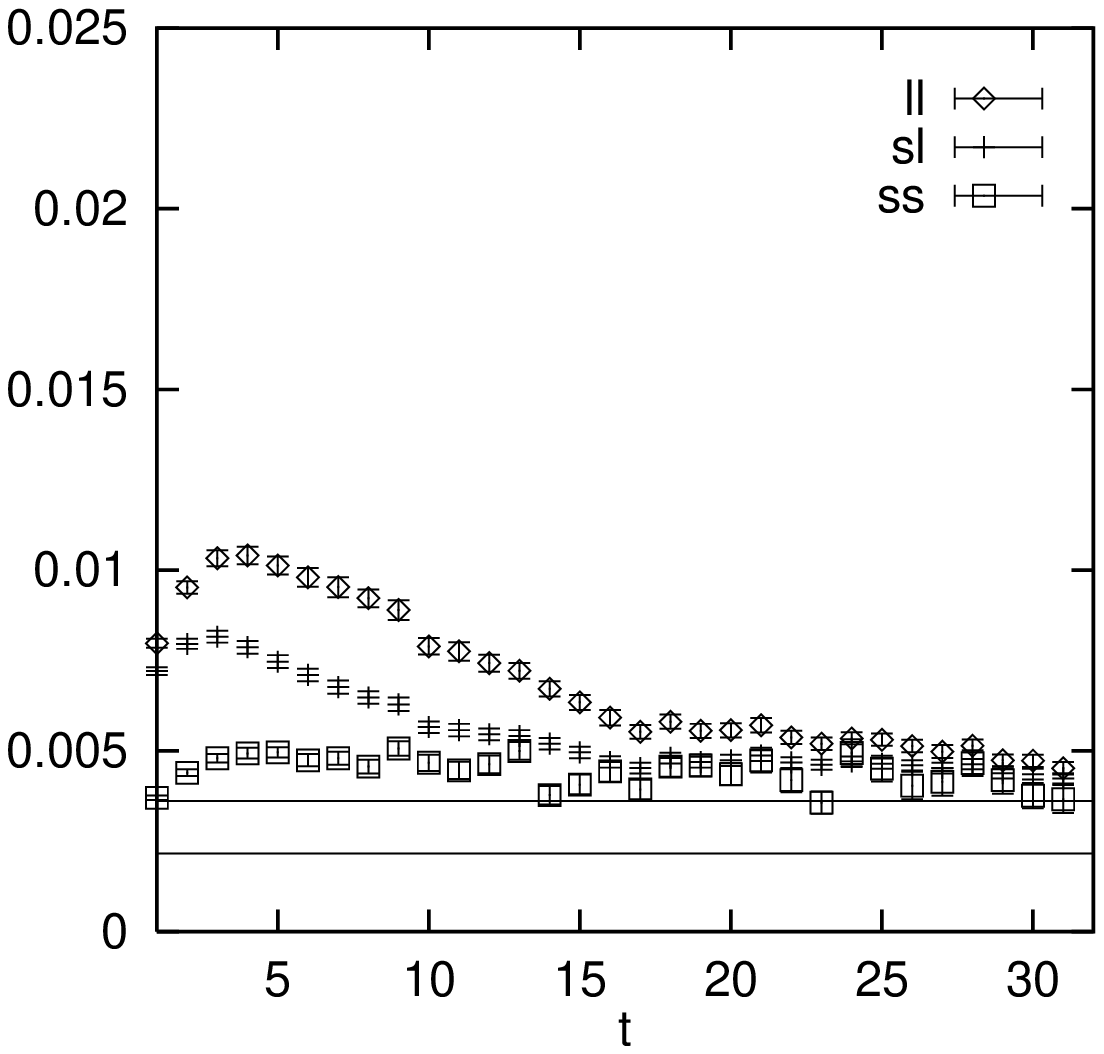,bbllx=71,bblly=52,
        bburx=388,bbury=346,width=0.481\linewidth}
     }
\hbox{
\epsfig{file=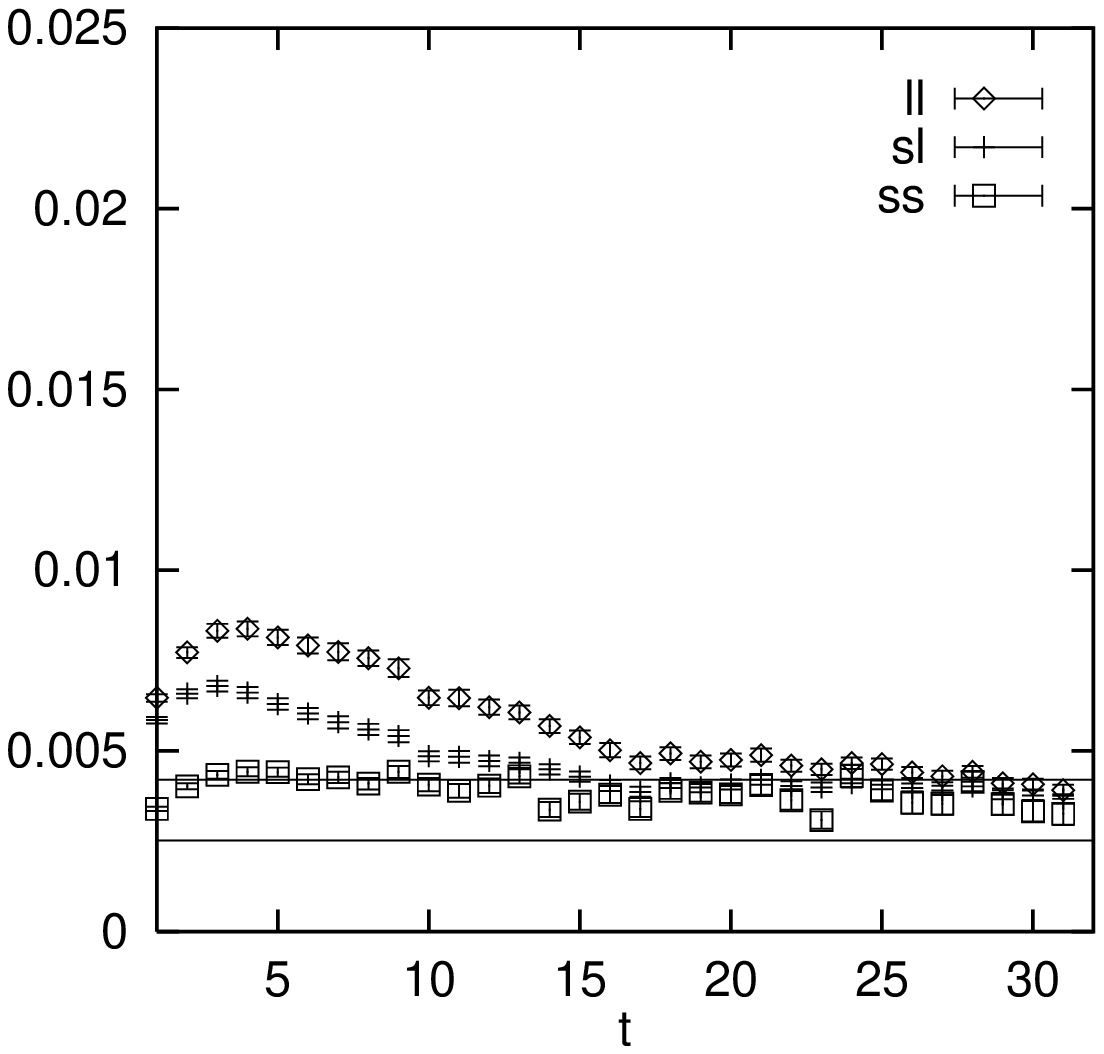,bbllx=71,bblly=52,
        bburx=388,bbury=346,width=0.481\linewidth}
\epsfig{file=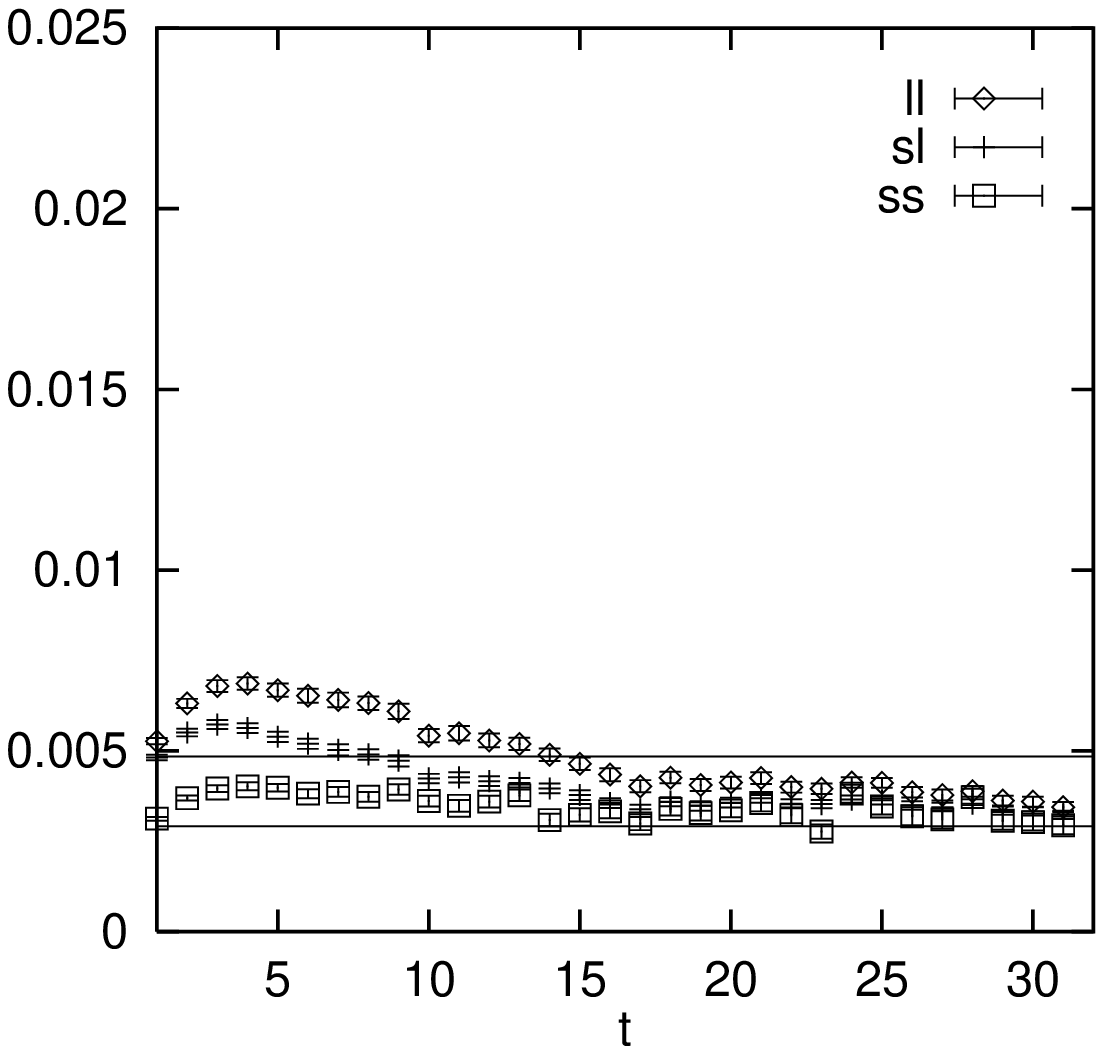,bbllx=71,bblly=52,
        bburx=388,bbury=346,width=0.481\linewidth}
     }
     }
\caption{Local masses $\ln{(G_m(t)/G_m(t+a_\tau))}$
         for the ${}^1S_0-{}^3S_1$ splitting in units of
         $a_\tau^{-1}$ from local-local (ll), smeared-local (sl) and
         smeared-smeared (ss) propagators 
         compared with the expectation obtained
         from $\Delta E=15-25$~MeV. We start with the lowest
         bare quark mass $a_\sigma M_Q=1.5$ in the upper left corner
         and go in intervals of 0.5 to the lower right corner
         with $a_\sigma M_Q=4.0$.}
\label{fig:hyp}
\end{figure}

The value of $a_\sigma^{-1}=\xi~M_{\eta_b}/(\xi M_{kin})$
is used to compare the expected $\Delta E_{hyp}$
with the measured ${}^3S_1-{}^1S_0$ splitting. From figs. \ref{fig:hyp}
we see that the best agreement is achieved for $M_Q=3.5$ which corresponds
to $a_\sigma^{-1} \approx 1.3$~GeV. There is a 15\% difference between
this value and the estimate from the string tension.
Both values are consistent within the expected accuracy.
A discrepancy of this size has also
been found in NRQCD studies at $T=0$ \cite{BSpec}.
A more stringent test and a 
better determination of $a_\sigma$ will be possible once the $1S-1P$
splitting has been measured \cite{FTNRQCD_long}.

The easiest quantities to calculate are $^1S_0$ propagators
for point sources. Scaled propagators, $H(T)=G(T)/G(T=0)$, show directly
the changes in the spectrum due to the temperature.
The zero point energy which
in principal can be computed using weak coupling perturbation theory is
the same for all three temperatures because we vary only $N_\tau$ keeping
all other parameters fixed.
The mass shift $\Delta M(T)=M_{meson}(T)-M_{meson}(T=0)$ can be extracted
from the slope of the scaled propagator at large time steps. A positive
slope indicates an increase while a negative slope
will be seen if masses decrease with temperature.
\begin{figure}[htb]
\vbox{
\hbox{
\epsfig{file=lr2_1_1.eps,bbllx=86,bblly=96,
        bburx=458,bbury=443,width=0.48\linewidth}
\epsfig{file=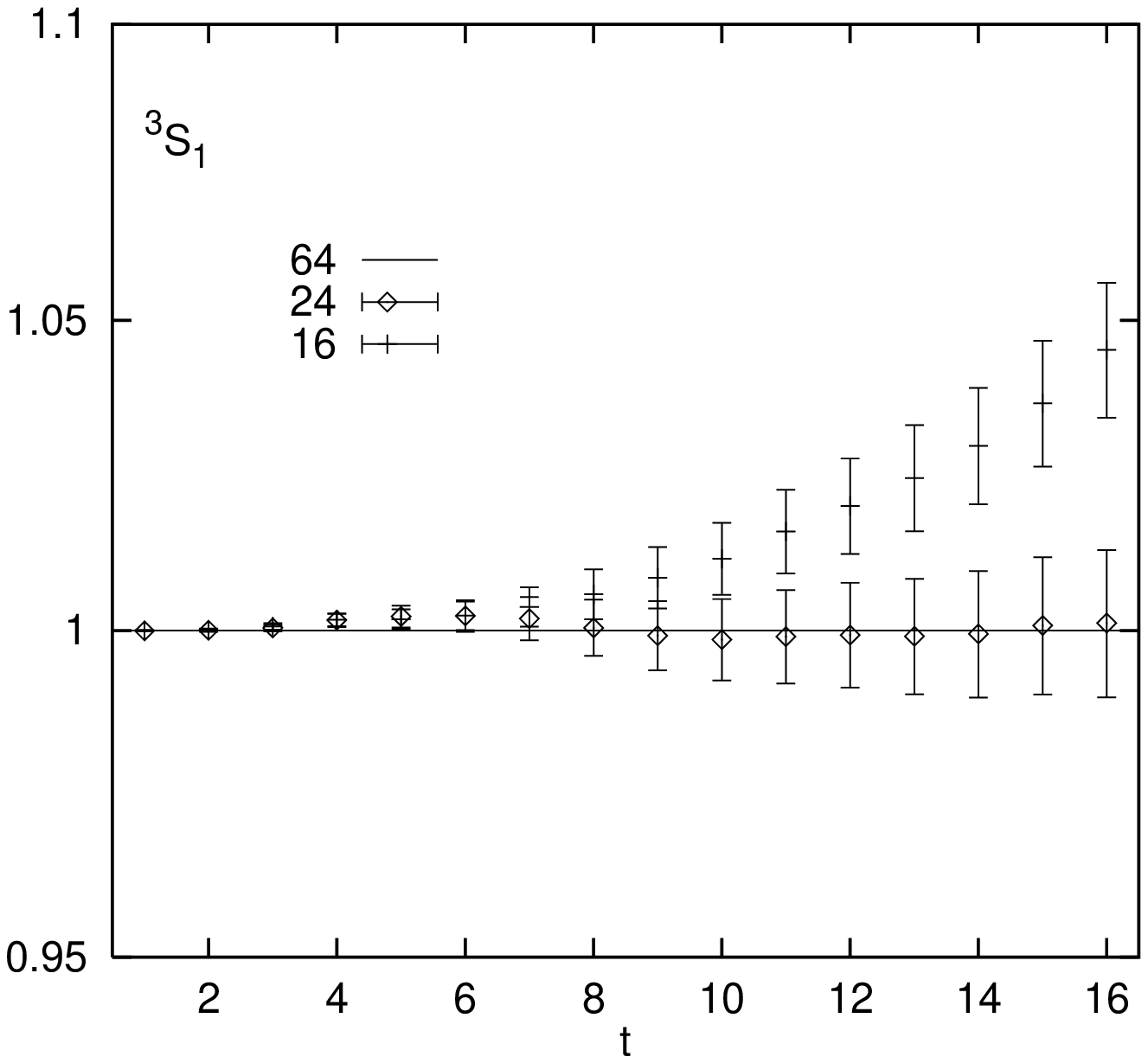,bbllx=88,bblly=50,
        bburx=460,bbury=397,width=0.48\linewidth}
     }
\hbox{
\epsfig{file=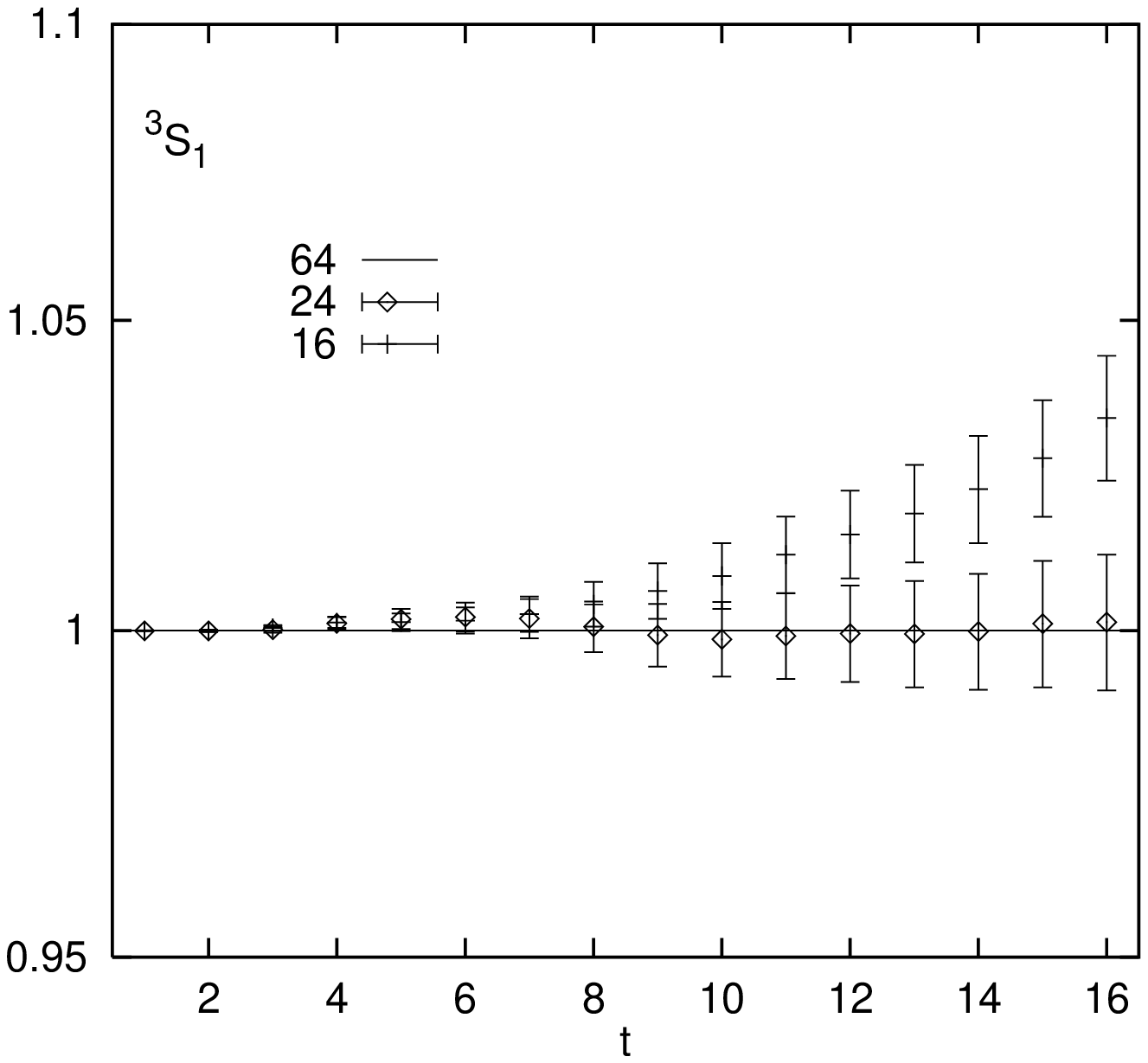,bbllx=88,bblly=50,
        bburx=460,bbury=397,width=0.48\linewidth}
\epsfig{file=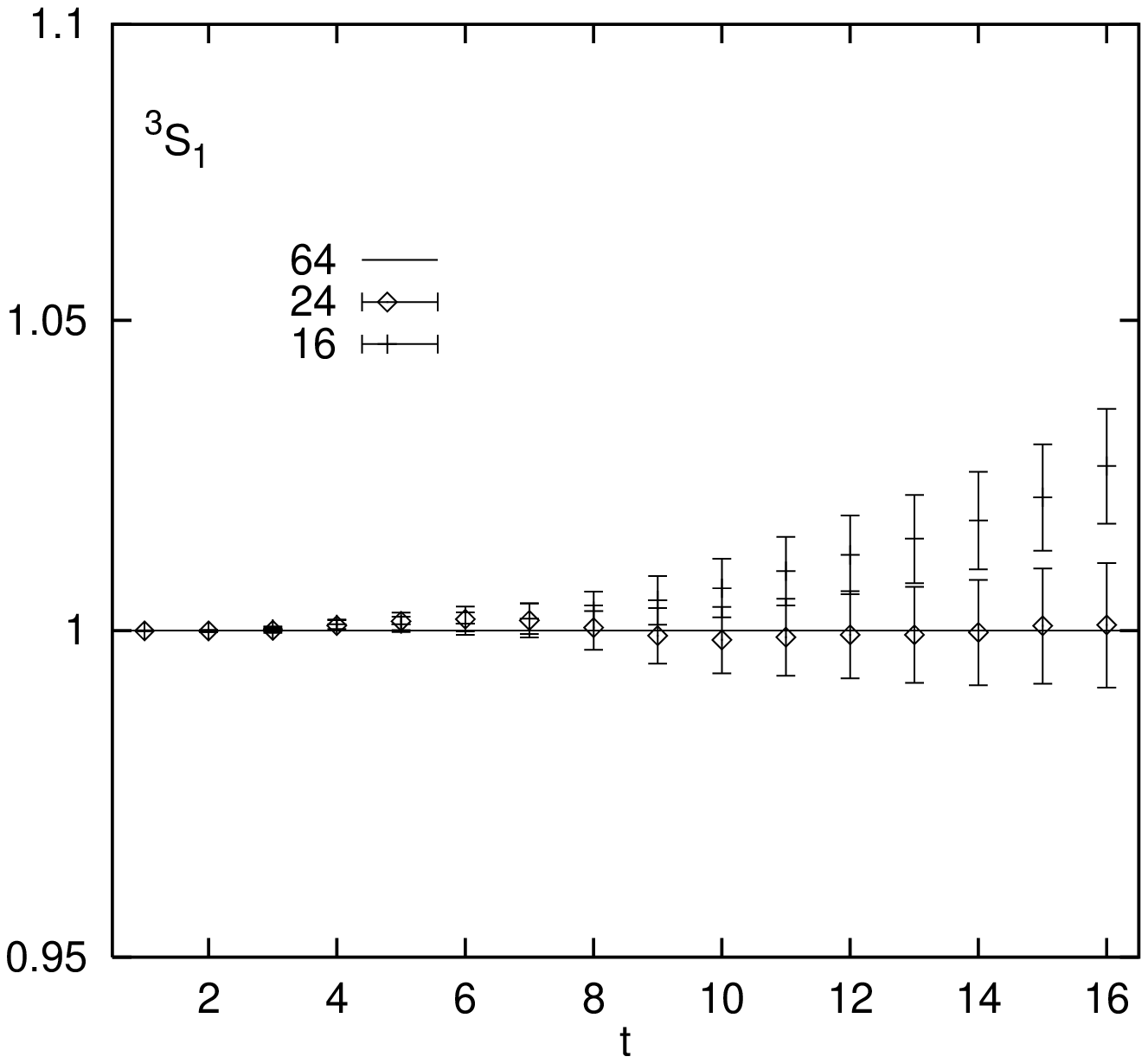,bbllx=88,bblly=50,
        bburx=460,bbury=397,width=0.48\linewidth}
     }
\hbox{
\epsfig{file=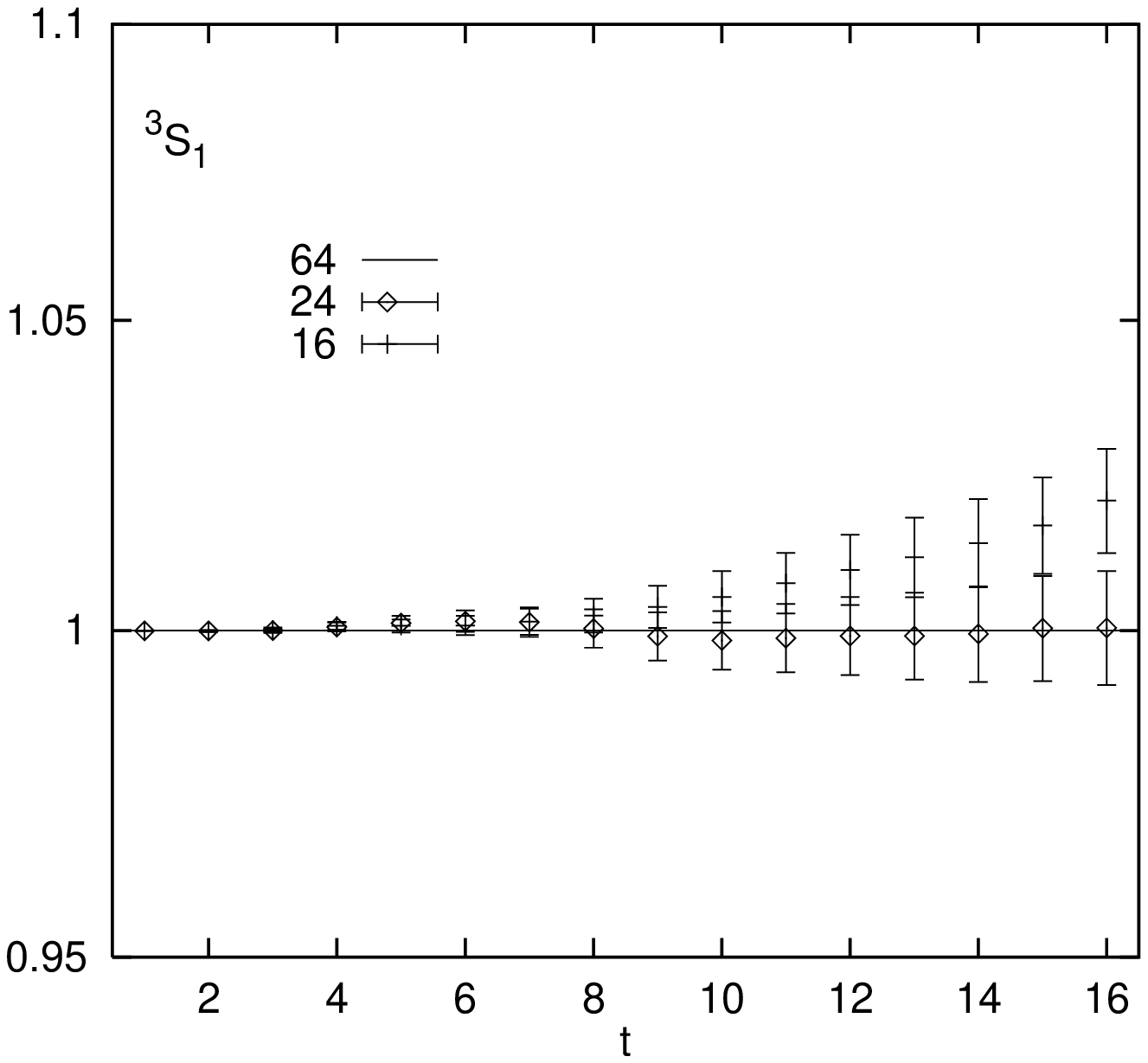,bbllx=88,bblly=50,
        bburx=460,bbury=397,width=0.48\linewidth}
\epsfig{file=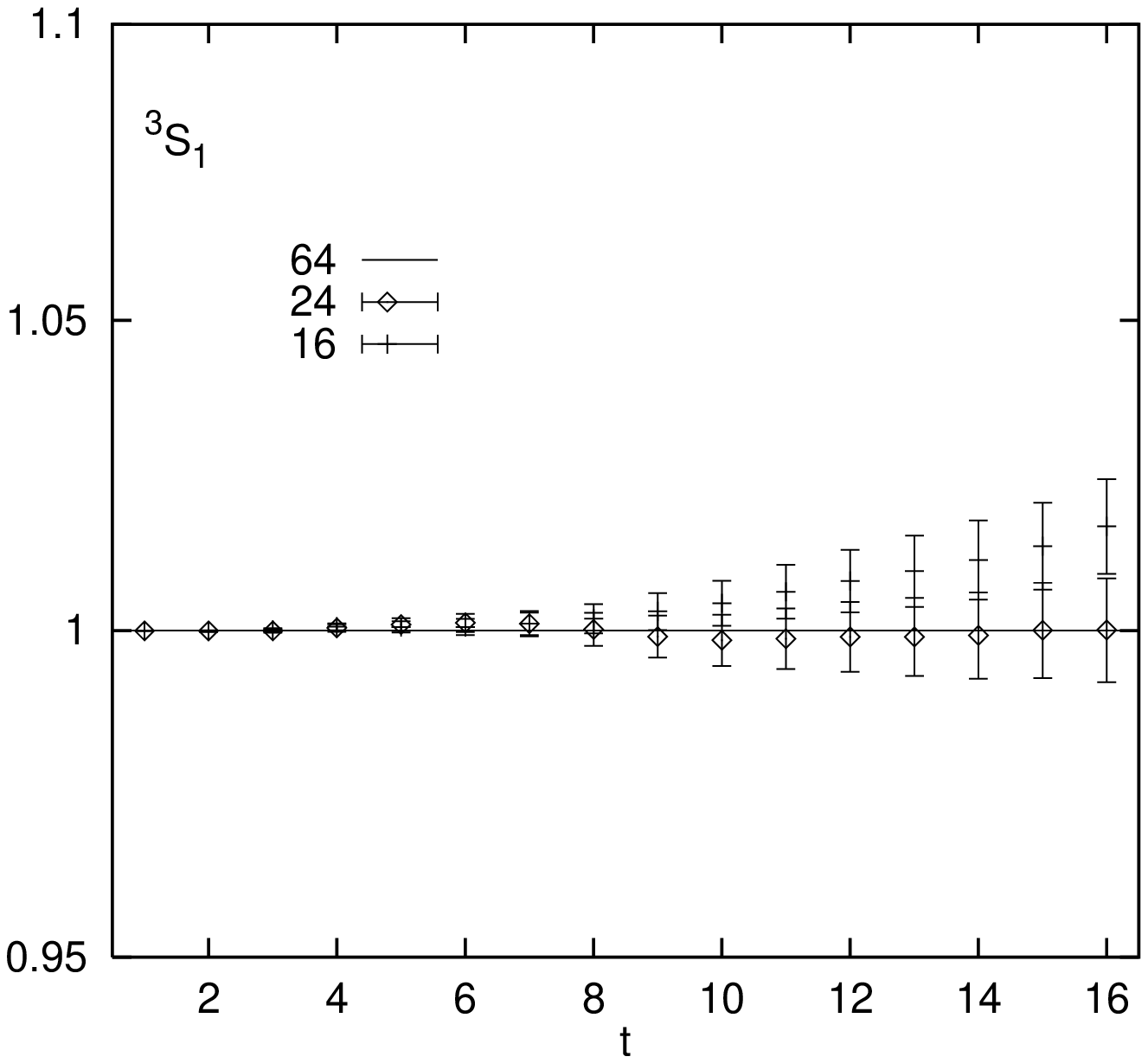,bbllx=88,bblly=50,
        bburx=460,bbury=397,width=0.48\linewidth}
     }
     }
\caption{Scaled local n=1~${}^3S_1$ meson propagators $H(T)$
         for 6 values of the bare quark mass starting 
         with the lightest mass in the upper left corner
         on a logarithmic scale.}
\label{fig:props}
\end{figure}

\begin{figure}[htb]
\vbox{
\hbox{
\epsfig{file=er2_1_1.eps,bbllx=89,bblly=24,
        bburx=460,bbury=370,clip=,width=0.48\linewidth}
\epsfig{file=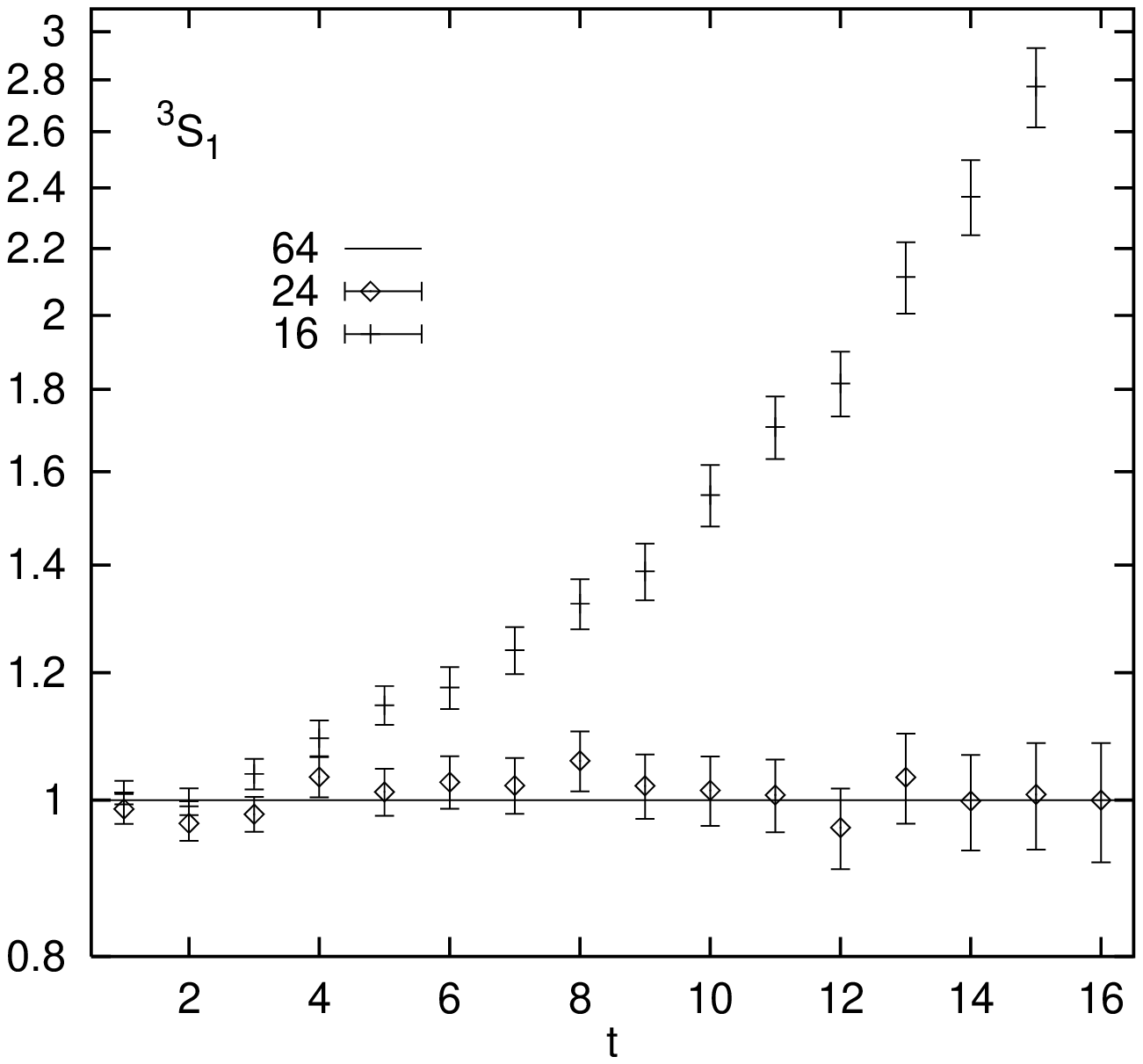,bbllx=88,bblly=50,
        bburx=460,bbury=397,width=0.48\linewidth}
     }
\hbox{
\epsfig{file=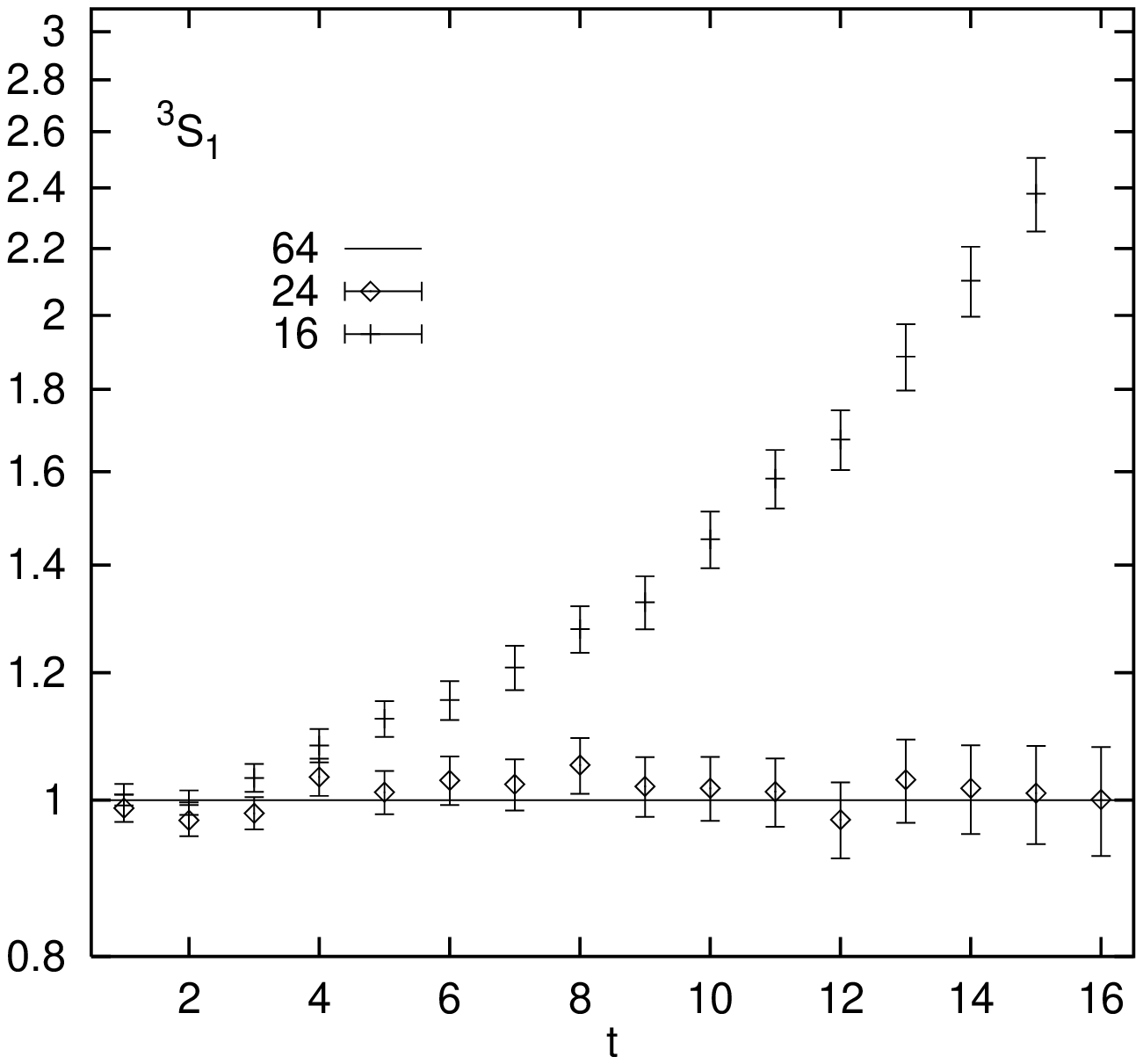,bbllx=88,bblly=50,
        bburx=460,bbury=397,width=0.48\linewidth}
\epsfig{file=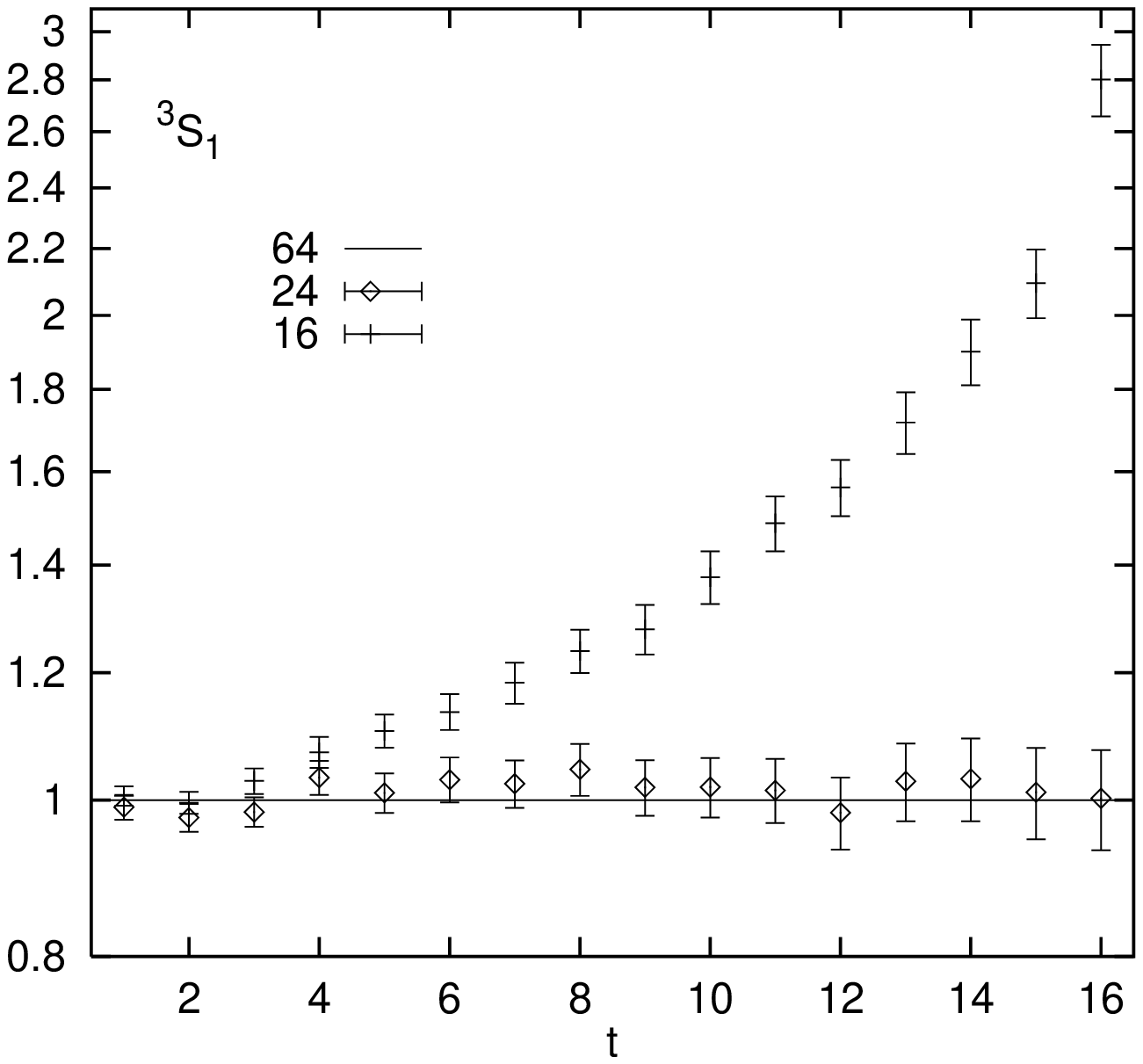,bbllx=88,bblly=50,
        bburx=460,bbury=397,width=0.48\linewidth}
     }
\hbox{
\epsfig{file=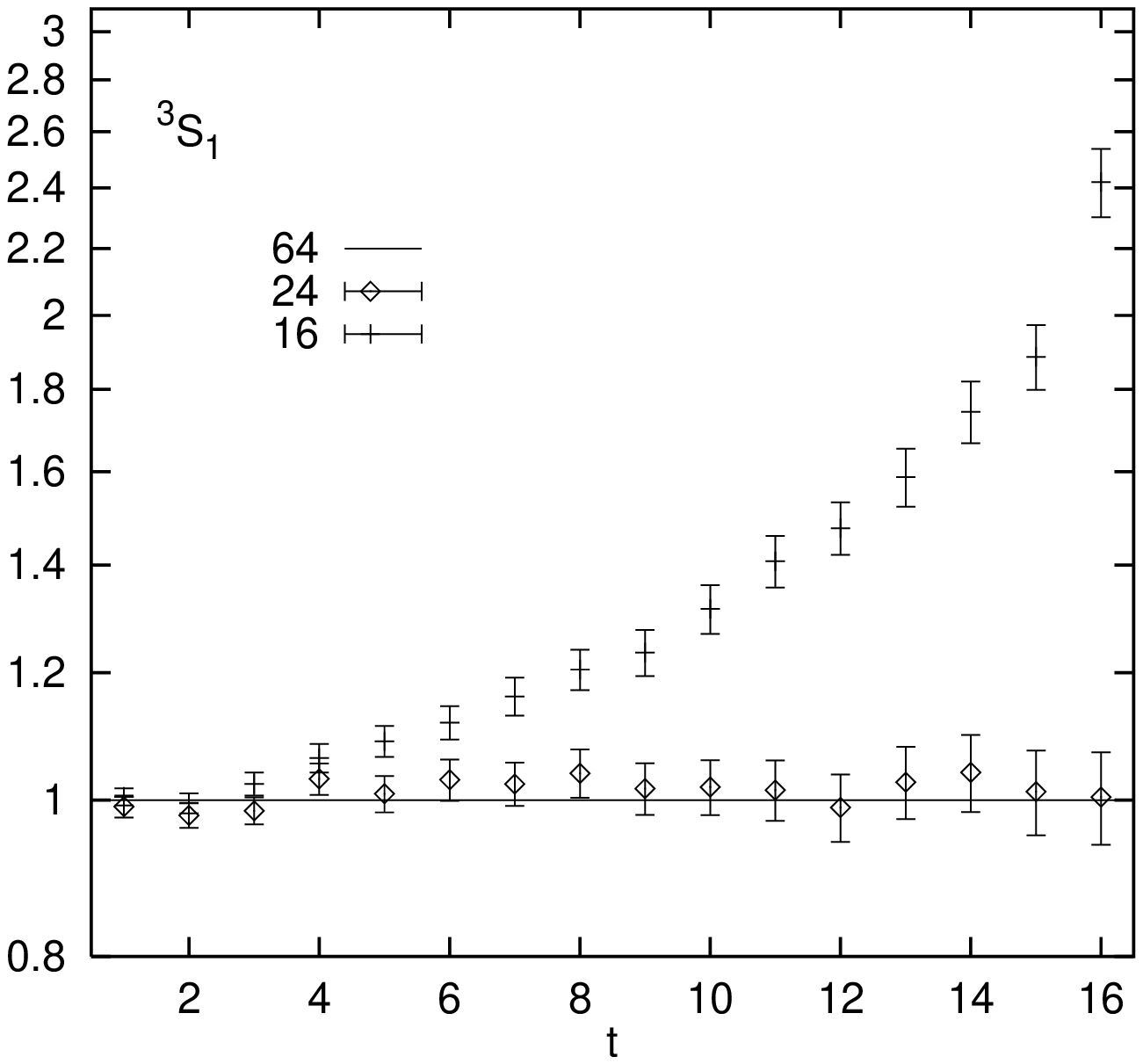,bbllx=88,bblly=50,
        bburx=460,bbury=397,width=0.48\linewidth}
\epsfig{file=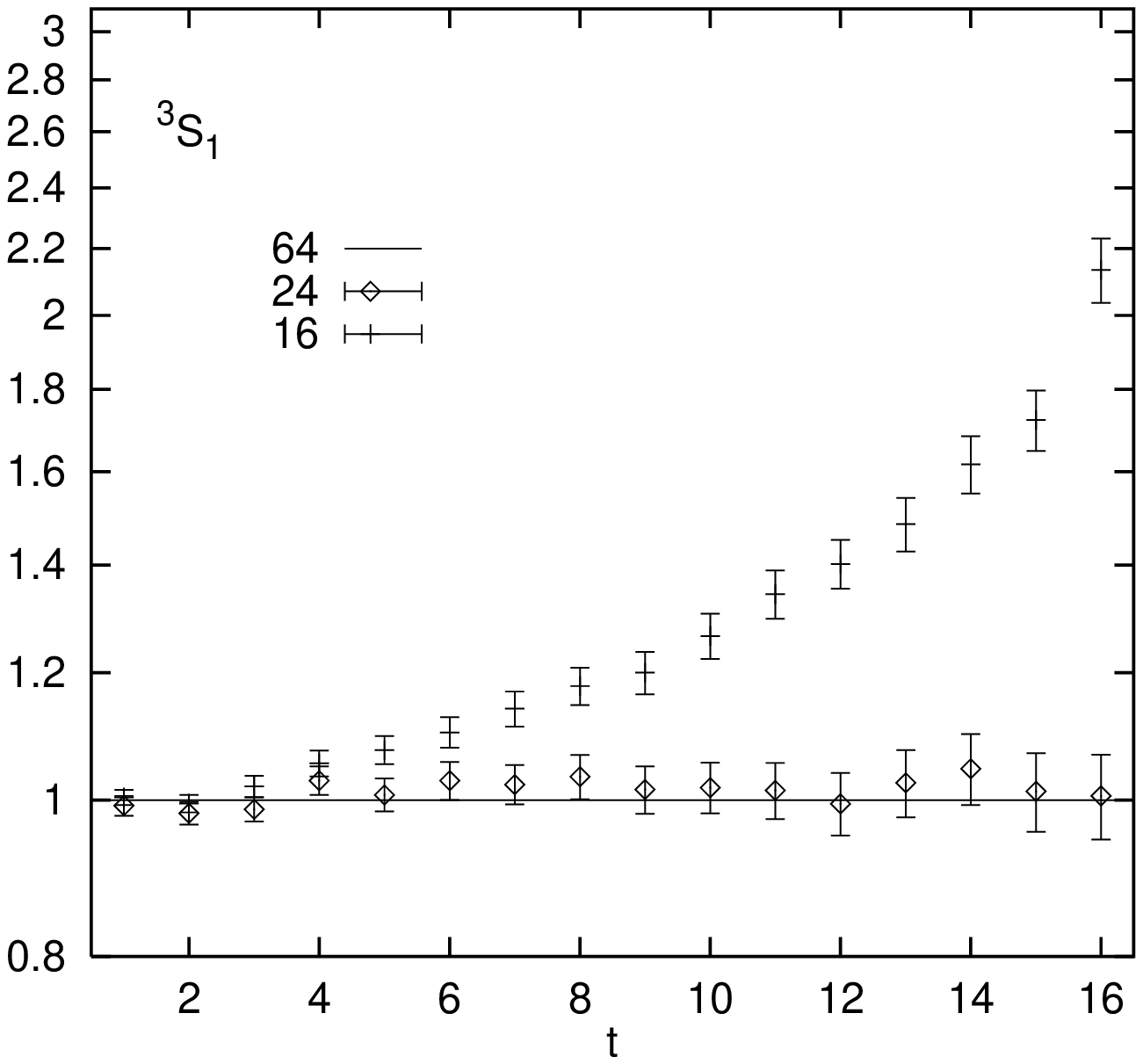,bbllx=88,bblly=50,
        bburx=460,bbury=397,width=0.48\linewidth}
     }
     }
\caption{Scaled smeared-smeared n=2~${}^3S_1$ meson propagators $H(T)$
         for 6 values of the bare quark mass increasing
         from upper left to lower right corner on a logarithmic scale.}
\label{fig:props_2n}
\end{figure}

Our numerical results show that
the signal for $Q\bar Q$ bound states persists to high temperature,
$T\approx 1.2~T_c$.
Compared to $T=0$ the ground state meson propagators change very little.
The observed temperature dependence of the scaled propagator 
(fig.~\ref{fig:props}) is weak.
Looking at it in more detail we see that the ground state meson mass
decreases with $T$. Fig.~\ref{fig:props} shows the scaled propagators
for $N_\tau=16$ and $24$. The scale is logarithmic and we expect
a linear form for an exponential behaviour, $G(t)\propto\exp{(-\Delta M~t)}$.
For $N_\tau=24$ which corresponds to $T \approx 0.8~T_c$ the scaled
propagator agrees with unity.
It does not show any significant change with temperature
in the entire mass range, $1.5 \le a_\sigma M_Q \le 4$.
For $N_\tau=16$ which corresponds to a temperature
 $T\approx 1.2~T_c$ we observe a clear
increase of the scaled propagator with Euclidean time.
This signals a broadening of the spectral function
and is an indication of a possible decrease of the effective
meson mass at this value of the temperature. As expected,
the effect becomes weaker as we go to larger quark mass.
To illustrate the order of magnitude we included a straight
line in fig.~\ref{fig:props} which corresponds to
a small shift of ~$-\Delta M=a_\tau^{-1} (\ln{1.06})/6\approx 12$~MeV. 
It is important to note that we used point sources and sinks
which do not distort the spectral function.
The observed effect is a signal that the mixture
of states excited by the local source gets broader and
possibly lighter with increasing temperature.
If there is no cancellation in the sense
that the ground state gets heavier while higher excitation become
lighter then particularly the ground state mass will decrease with temperature.

It is expected that first excited states dissociate immediately
above the phase transition. In fig.~\ref{fig:props_2n} we see no effect
at $T\approx 0.8~T_c$ but a strong increase of the propagators
at $T\approx 1.2~T_c$. Compared to the ground state the
increase is much stronger so that it seems likely that the first
excited state is dissociated at this temperature.
Again, the order of magnitude of the effect is illustrated
in fig.~\ref{fig:props_2n} by a straight line which now 
corresponds to a shift of ~$-\Delta M=a_\tau^{-1} (\ln{3})/10\approx 240$~MeV.
Since there is no local operator projecting on the first excited state
this result has been obtained with a smeared source and sink. The trial 
wave-function can only approximate the temperature wave-function causing a
possible residual distortion of the quarkonium propagator. To overcome
this problem a measurement of the temperature wave function is
desirable \cite{FTNRQCD_long}.

Our numerical results can be contrasted with an expectation from the density
matrix formalism. If we assume that the thermal meson state consists
of a mixture of zero temperature states with a statistical
weight given by the Boltzmann factor
 $G(t,T)=\sum_i \exp{(-M_i/T)} \exp{(-M_i~t)}$ then we would expect
a decrease of the scaled propagator with $t$ because more and more heavier
states contribute as the temperature is increased.
The observed increase of the scaled propagators cannot be
explained by a simple temperature dependent admixture of higher mesonic
states with negligible spectral width.

\section{Summary}
We have performed a first non-perturbative
study of the heavy meson spectrum and dispersion
relation at finite temperature using FT-NRQCD. 
We found no change in the spectrum for temperatures
below the deconfinement transition
and strong non-trivial effects above $T_c$.

A comparison of S-state meson propagators
at $T=0$ and $T\approx 0.8~T_c$ showed no significant
differences in the entire Euclidean time range, $t\le 16$.
The local operator used for the ground state does not distort or re-weight
the spectral function. At low time steps it gets contributions not only
from the ground state but also from higher excitations. If this quantity
is unchanged at large $t$ so is the true ground state.

A further increase of the temperature to $T \approx 1.2~T_c$
showed a clear signal for a broadening of the spectral function
of the ground state. This can be interpreted as an indication
for a small decrease of the meson mass with temperature.
The phenomenon is stronger the lighter the quark mass is.
As expected we see that the first excited S-state is more 
susceptible to the temperature. Across the deconfinement transition
we observe a strong change between $T\approx 0.8~T_c$ and $T\approx 1.2~T_c$.
This can be taken as a first indication for the dissociation of these states.

Qualitatively all our observations
are in accord with the expectations from the phenomenological picture
of Debye-screening in a potential model.
The heavier the mesons the smaller they are and the less they
feel the screening. This has important implications for the
deconfinement transition. The confined degrees of freedom are not released
at a single temperature $T_c$ but the transition to a quark gluon plasma
proceeds in steps. The determination of the dissociation
temperature of different bound states of heavy quarks
is an interesting problem which deserves further efforts \cite{FTNRQCD_long}.

There is a discrepancy with other models that predict a strong effect
already below $T\le T_c$ such as a weakening of $\sigma(T)$.
The string tension surely has to vanish at $T=T_c$. However, this effect
is a long range phenomenon. At smaller distances remaining strong
interactions can still give rise to an effective string
tension $\sigma(R,T)>0$.
Furthermore, we see only a small change of the ground state mass
up to $T=1.2~T_c$ unlike predicted by most phenomenological models.
A crossover between a behaviour dominated
by the strong coupling constant at small distances and the
string tension at larger separations is not observed
in the accessible mass range. This gives further support to the observation
that the perturbative form of the heavy quark potential is wrong in
the temperature range under consideration.

A precise measurement of the shift of
the meson mass at higher temperature requires further work \cite{FTNRQCD_long}.
For the future we intend to measure the wave-function and use it to
improve the ground-state overlap of our operators. We expect that
a multi-state fit with the improved operators allows a determination
of the spectrum at higher temperature.

\section{Acknowledgments}
We thank Frithjof Karsch, Helmut Satz and
Edwin Laermann for useful conversations and the
SESAM-collaboration for continuous support.
The computations were done on the Quadrics Q4 and the
Connection Machine CM2 in Wuppertal.
J.F. thankfully acknowledges a fellowship from the
Deutsche Forschungsgemeinschaft.


\begin{thebibliography}{99}

\bibitem{MatsuiSatz}
T.Matsui and H.Satz, Phys. Lett. B178 (1986) 416.

\bibitem{qprod}
A.~Sansoni et al. (CDF Collaboration), 
{\em Quarkonia Production at CDF}, FERMILAB-CONF-96-221-E;
E.~Braaten, Nucl. Phys. A610 (1996) 386c;
M.L.~Mangano, CERN-TH/95-190;
M.L.~Mangano and A.~Petrelli, e-Print Archive: hep-ph/9610364.

\bibitem{pot}
T.~Hashimoto, O.~Miyamnura, K.~Hirose and T.~Kanki,
Phys. Rev. Lett. 57 (1986) 2123.

\bibitem{pot1}
T.~Hashimoto, K.~Hirose, T.~Kanki, O.~Miyamura , Z. Phys. C38 (1988) 251. 

\bibitem{precursory}
S.~Hioki, T.~Kanki and O.~Miyamura, Prog. Theor. Phys. 85 (1991) 603.

\bibitem{sum_rules}
R.J.~Furnstahl, T.~Hatsuda and S.H.~Lee, Phys. Rev. D42 (1990) 1744.

\bibitem{NJL}
F.O. Gottfried and S.P. Klevansky, Phys. Lett. B286 (1992) 221.

\bibitem{NaNu}
F.S.~Navarra and C.A.A.~Nunes,
Phys. Lett. B356 (1995) 439.

\bibitem{ftmes93}
T.~Hashimoto, A.~Nakamura and I.O.~Stamatescu,
Nucl. Phys. B400 (1993) 267; Nucl. Phys. B406 (1993) 325.

\bibitem{ftmes95}
K.~Akemi et al. (QCD-Taro collaboration),
Nucl. Phys. B (Proc. Suppl.) 42 (1995) 445;
M.~Fujisaki et al. (QCD-Taro collaboration),
Nucl. Phys. B (Proc. Suppl.) 53 (1997) 426,

\bibitem{CTHD94}
C.T.H.~Davies et al., Phys. Rev. D50 (1994) 6963.

\bibitem{CTHD95}
C.T.H.~Davies et al., Phys. Rev. D52 (1995) 6519.

\bibitem{MaPo87}
E. Manousakis and J. Polonyi, Phys. Rev. Lett. 58 (1987) 847.

\bibitem{SST93}
G.~Bali, J.~Fingberg, U.M.~Heller, F.~Karsch and K.~Schilling,
Phys. Rev. Lett. 71 (1993) 3059.

\bibitem{KLL95}
F.~Karsch, E.~Laermann and M.~L\"utgemeier,
Phys. Lett. B346 (1995) 94.

\bibitem{PDG}
R.M.~Barnett et al. (Particle Data Group), Phys. Rev. D54 (1996) 1.

\bibitem{Tdiss}
F.~Karsch and H.~Satz, Z. Phys. C51 (1991) 209.

\bibitem{Lepage}
G.P.~Lepage and B.A.~Thacker, Phys. Rev. D43 (1991) 196;
K.~Hornbostel, G.P.~Lepage, L.~Magnea, U.~Magnea and C.~Nakhleh,
Phys. Rev. D46 (1992) 4052.

\bibitem{asym}
F.~Karsch, Nucl. Phys. B205 (1982) 285.

\bibitem{aniso}
G.~Burgers, F.~Karsch, A.~Nakamura and I.O.~Stamatescu,
Nucl. Phys. B304 (1988) 587.

\bibitem{colin}
C.~Morningstar, Nucl. Phys. Proc. Suppl. 53 (1997) 914;
e-Print Archive: hep-lat/9704011.

\bibitem{Montvay}
H.~Joos and I.~Montvay, Nucl. Phys. B225 (1983) 565.

\bibitem{Satz}
F.~Karsch, M.T.~Mehr and H.~Satz, Z. Phys. C37 (1988) 617.

\bibitem{Edwin}
E.~Laermann, Nucl. Phys. B (Proc. Suppl.) 42 (1995) 120.

\bibitem{Bengt}
B.~Petterson, Nucl. Phys. A525 (1991) 237.

\bibitem{columbia}
L.I.~Unger, Phys. Rev. D48 (1993) 3319.

\bibitem{Henning}
P.A.~Henning, Phys. Rep. 253 (1995) 235.

\bibitem{tia}
M.~Garc\'ia P\'erez, J.~Snippe and P.~van~Baal,
{\it Testing Improved Actions},
e-Print Archive: hep-lat/9607007.

\bibitem{anisotrop}
M.~Garc\'ia P\'erez and P.~van~Baal,
Phys. Lett. B392 (1997) 163.

\bibitem{rad}
C.~Morningstar, Phys. Rev. D50 (1994) 5902.

\bibitem{BF}
G.~Bali, K.~Schilling and A.~Wachter,
{\it Complete O$(v^2)$ corrections to the static
     interquark potential from SU(3) gauge theory},
e-Print Archive: hep-lat/9703019.

\bibitem{Cella94}
G.~Cella, G.~Gurci, A.~Vicere and B.~Vigna, Phys. Lett. B333 (1994) 457.

\bibitem{thermo}
G.~Boyd, J.~Engels, F.~Karsch, E.~Laermann, C.~Legeland,
M.~L\"utgemeier and B.~Petersson, Nucl. Phys. B469 (1996) 419.

\bibitem{Tsukuba}
C.T.H.~Davies,
{\it The Spectrum from Lattice NRQCD},
e-Print Archive: hep-lat/9705039.

\bibitem{BSpec}
S.~Collins et al., Phys. Rev. D54 (1996) 5777.

\bibitem{FTNRQCD_long}
J.~Fingberg, work in progress.

\end{thebibliography}
\end{document}